\documentclass[aps,twocolumn,groupedaddress,showpacs]{revtex4}
\usepackage{amssymb,amsmath,epsfig}
\usepackage{graphicx}
\usepackage{subfigure}

\newcommand{\ignore}[1]{}

\begin{document}

\title{Controlling Effect of Geometrically Defined Local Structural Changes on Chaotic Hamiltonian Systems}


\author{Yossi Ben Zion$^{1}$ and Lawrence Horwitz$^{1,2,3}$}
 \affiliation{$^1$Department of
Physics, Bar Ilan University, Ramat
Gan 52900, Israel\\
$^2$Department of Physics, Ariel University Center of Samaria, Ariel 40700, Israel\\
 $^3$School of Physics, Tel Aviv University, Ramat Aviv
69978, Israel}

\begin{abstract}

An effective characterization of chaotic conservative Hamiltonian
systems in terms of the curvature associated with a Riemannian
metric tensor derived from the structure of the Hamiltonian has been
extended to a wide class of potential models of standard form
through definition of a conformal metric.  The geodesic equations
reproduce the Hamilton equations of the original potential model
through an inverse map in the tangent space. The second covariant
derivative of the geodesic deviation in
 this space generates a dynamical curvature, resulting in
(energy dependent) criteria for unstable behavior different from the
usual Lyapunov criteria. We show here that this criterion can be
constructively used to modify locally the potential of a chaotic
Hamiltonian model in such a way that stable motion is achieved.
Since our criterion for instability is local in coordinate space,
these results provide a new and minimal method for achieving control
of a chaotic system.

\end{abstract}

\maketitle
\bigskip
\bigskip

\section{Introduction}

\par The issue of controlling chaos has attracted great interest in the
last two decades; much work has been done in the case of dissipative
systems. Ott, Grebogi and York (OGY)~\cite{ott}developed a method by
which chaos can be suppressed by making small time-dependent
perturbations in order to shadow one of the infinitely many periodic
orbits embedded in the chaotic attractor. This method has been
extended by many following
studies~\cite{kapitaniak,dressler,pyragas} and successfully applied
to experimental systems~\cite{ditto}.

As a result of the absence of attractors in conservative systems and
the complex nature of the phase space combining regular and
irregular regions, constructing an effective method has remained a
challenging question. Lai Ding and Grebogi (LDG) extended the OGY
method to Hamiltonian systems by incorporating the notion of stable
and unstable directions at each periodic point.

In relation to the above question, Chandre \emph{et
al.}~\cite{chandre,chandre1,vittot} have developed a method for
constructing barriers in phase space to subdue chaotic behavior for
large times. This method is based on the introduction of a
specifically designed small control term which changes the dynamics
from chaotic to regular behavior.  Zhang \emph{et al.}~\cite{zhang}
have introduced a method for controlling chaos in two-dimensional
Hamiltonian system which they called adaptive integrable mode
coupling based on the separation of the system into two coupled
subsystems one of which is stable and the other unstable; when the
unstable system comes into the vicinity of the integrable system the
conditions are reset resulting in an effective adaptive control.
Cartwright \emph{et al.}~\cite{cartwright} have constructed a method
which permits stabilization of KAM islands through forward iteration
of the orbits, and transforming them into global attractors of the
embedded system. Ciraolo \emph{et al.}~\cite{ciraolo} have used a
method of adding a small perturbation to modify systems, such as
nonlinear plasmas, to follow more regular motion. Additional
efficient methods have been proposed
in~\cite{Kulp,bolotin,ding,yugui,zhihua}; some of these methods
require tracking the trajectories while others involve the addition
of interacting terms for which the criteria are somewhat sensitive.

Geometric approaches for the analysis of the stability of a given
Hamiltonian have been widely
discussed~\cite{jacobi,hadamard,casettiReport,casettiPRE,cipriani,BenZionPRL,BenZionPRE1,BenZionPRE2,Kawabe,szyd1,szyd2,szyd3,szyd4},
for which curvature of a manifold is associated with stability.

The pioneering work of Oloumi and Teychenn\'e~\cite{oloumi} proposes
a stabilization method by considering the Gaussian curvature of the
potential energy as a source of chaos. Even if the condition of
negative curvature of the potential and instability of the dynamics
are not completely equivalent they were able to successfully control
the instability by avoidance of negative curvature regions of the
potential energy (ANCRP).

In this work we make use of a recently developed geometrical
criterion~\cite{BenZionPRL} which is highly sensitive to instability
in Hamiltonian systems, and has been shown to be in complete
agreement with the results of the numerical technique of surface of
section (Poincar\'e plot) and more effective than other geometric
methods ~\cite{BenZionPRE2}. This criterion is based on an
equivalence between motions generated by Hamiltonian in the standard
form with quadratic kinematic terms and additive potentials (which
we shall call the {\it Hamilton} description) and a Hamiltonian in
which the dynamics is described by a metric-type function of
coordinates multiplying the momenta in bilinear form (which we shall
call the {\it Gutzwiller}
description)~\cite{gutzwiller,curtis,moser,eisenhardt,arnold}. The
mapping between these equivalent descriptions was first introduced
by Appel~\cite{appell,cartan,landau,zerzion}. The criterion for
stability developed in ~\cite{BenZionPRL} establishes an inverse
mapping of the motion described completely geometrically in the
Gutzwiller space back to the Hamilton description, carrying with it
the covariance under diffeomorphisms that is a property of the
Gutwiller dynamics. The resulting orbits in the Hamilton
description, with the special choice of coordinates serving as the
basis for the Appel type relation, reduce to the standard
description of the Hamilton orbits under the Hamilton equations.  In
general geometric form, as geodesic equations, however, they are
subject to analysis in terms of geodesic deviation, and the
resulting formula can be reduced to a computation in the special
coordinates of a new (symmetric) matrix valued criterion for
stability~\cite{BenZionPRL} which has been shown to be very
effective in a wide range of
examples~\cite{BenZionPRE1,BenZionPRE2,yahalom}.
\par In the analysis of these systems, it was shown that the presence
of negative eigenvalues for the stability matrix in the admissible
physical region (for which $E>V$) results in a chaotic type
Poincar\'e plot.  Since the condition is local in coordinate space,
the result necessarily implies some degree of ergodicity. There has,
however, been no proof that the existence of negative eigenvalues is
not strongly associated with the structure of the dynamics more
globally, as is often the case, for example, for analytic function
theory, where the presence of a complex pole implies a distortion of
a relatively large region of the complex plane.  The results of our
application to the control of dynamical systems, however, indicates
that the regions of negative eigenvalue correspond to a very
localized phenomena.  The observed instability they induce appears
to be due to the passage of orbits through these regions, and they
are not strongly correlated with the more global structure of the
dynamics.
\par What we have done here is to identify the regions of negative
eigenvalue in the coordinate space, and modify the Hamiltonian in
these regions locally, removing the source of instability either by
removing the instability inducing terms or varying the coupling to
these terms to bring the Hamiltonian closer to an integrable form in
these regions. The remaining part of the space is left to develop
according to the full, nonlinear, symmetry breaking, evolution.  The
effect on the Poincar\'e plots, as we show here, is dramatic.  This
result provides strong evidence that chaotic Hamiltonian systems
derive their properties from sources that may be thought of as
highly localized in the regions of negative eigenvalues for the
stability matrix.  In addition to giving an interesting insight into
the nature of Hamiltonian chaos, it also provides an effective
control method in which the Hamiltonian dynamics is left to exercise
its full function in a large domain of configurations, but is only
constrained in prescribed local regions of the configuration space.
\par In the next section, we review the basic ideas underlying the
geometrical criterion, and in the following section we give
numerical results for the application of our control procedure for
the cases of an oscillator with broken symmetry and a potential
which may be derived from the Toda form.

\begin{figure*}
  \centering
   $(a_{I})$ \subfigure{ \includegraphics[width=0.25\textwidth]{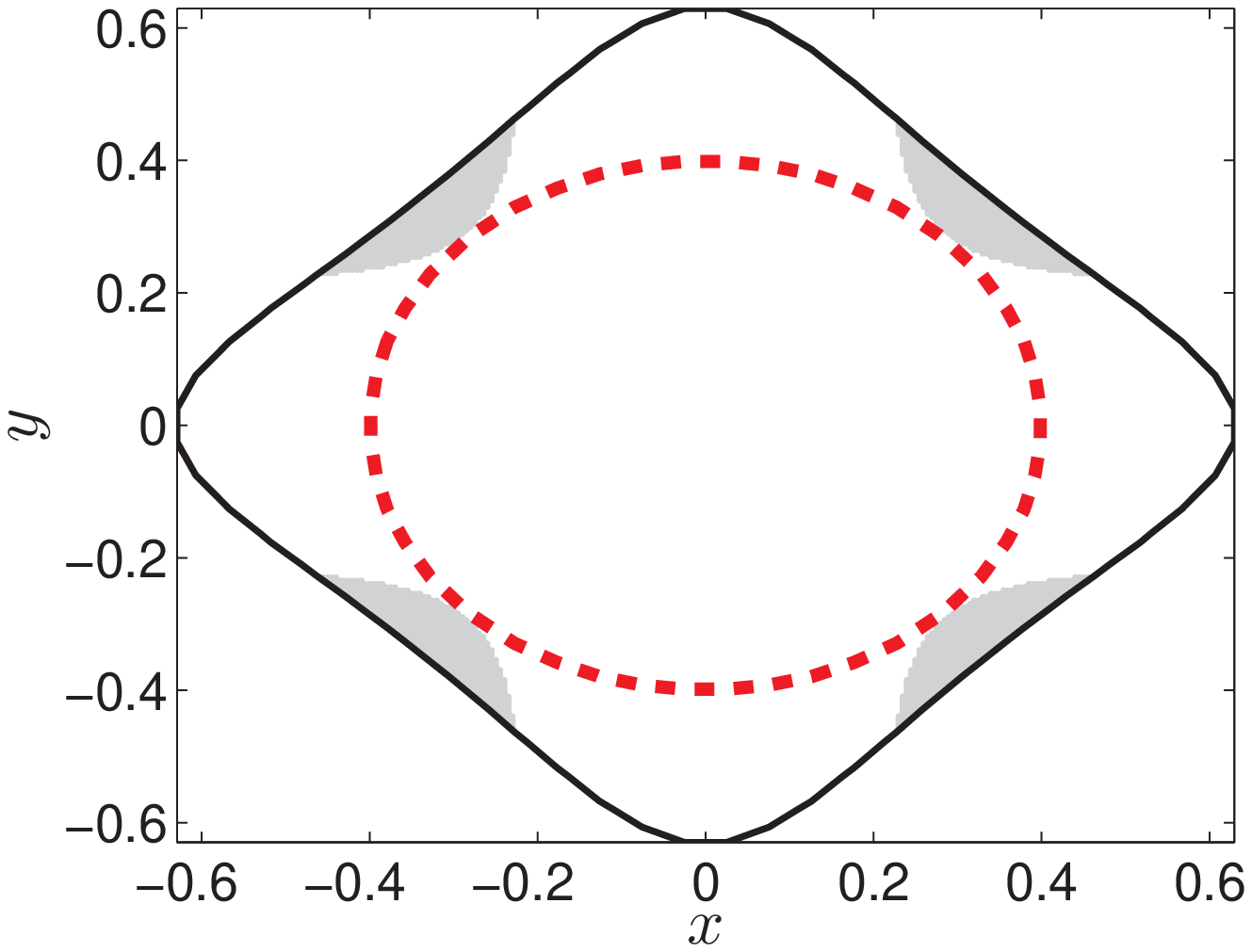}}
   $(a_{II})$\subfigure{\includegraphics[width=0.25\textwidth]{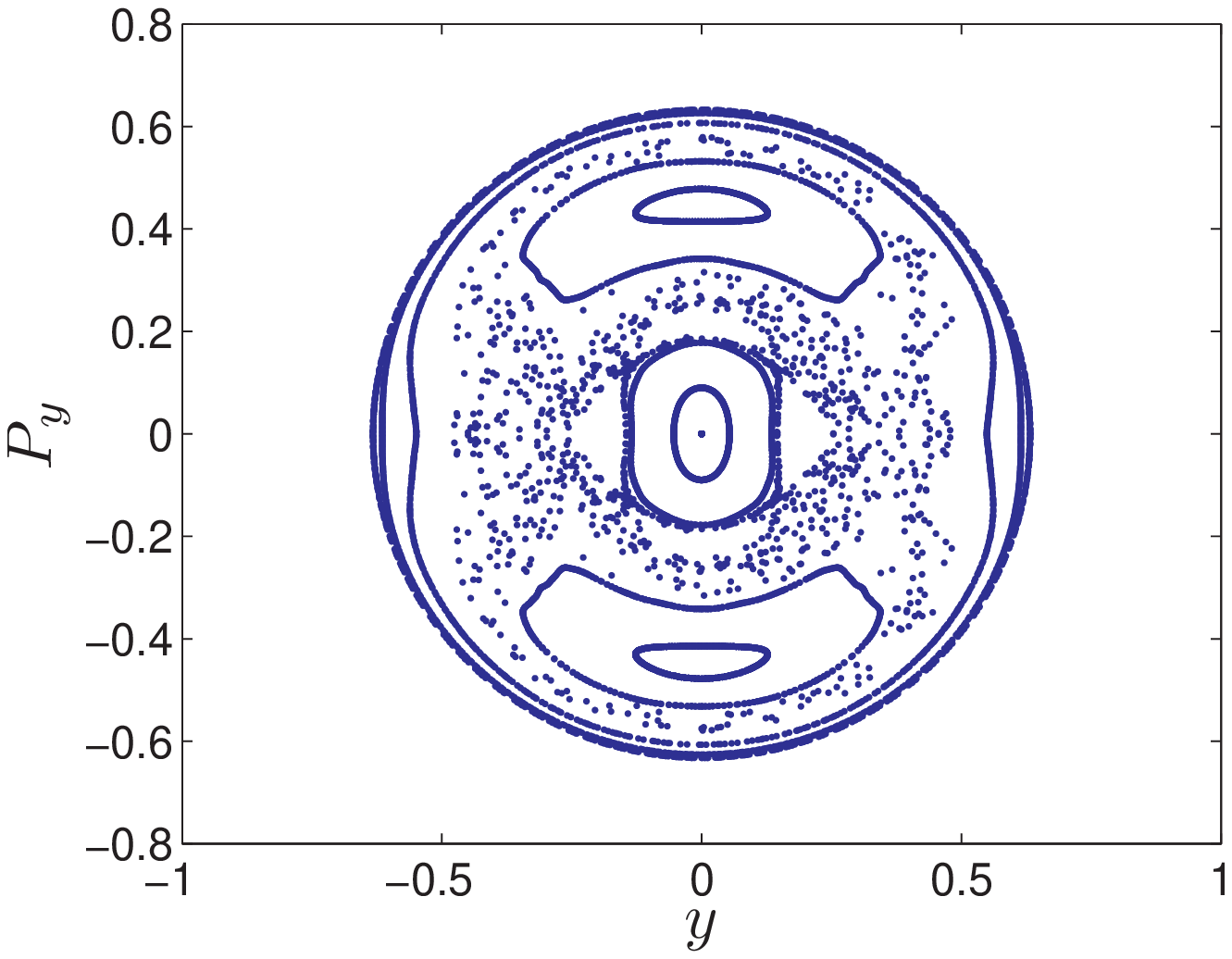}}
  $(a_{III})$\subfigure{\includegraphics[width=0.25\textwidth]{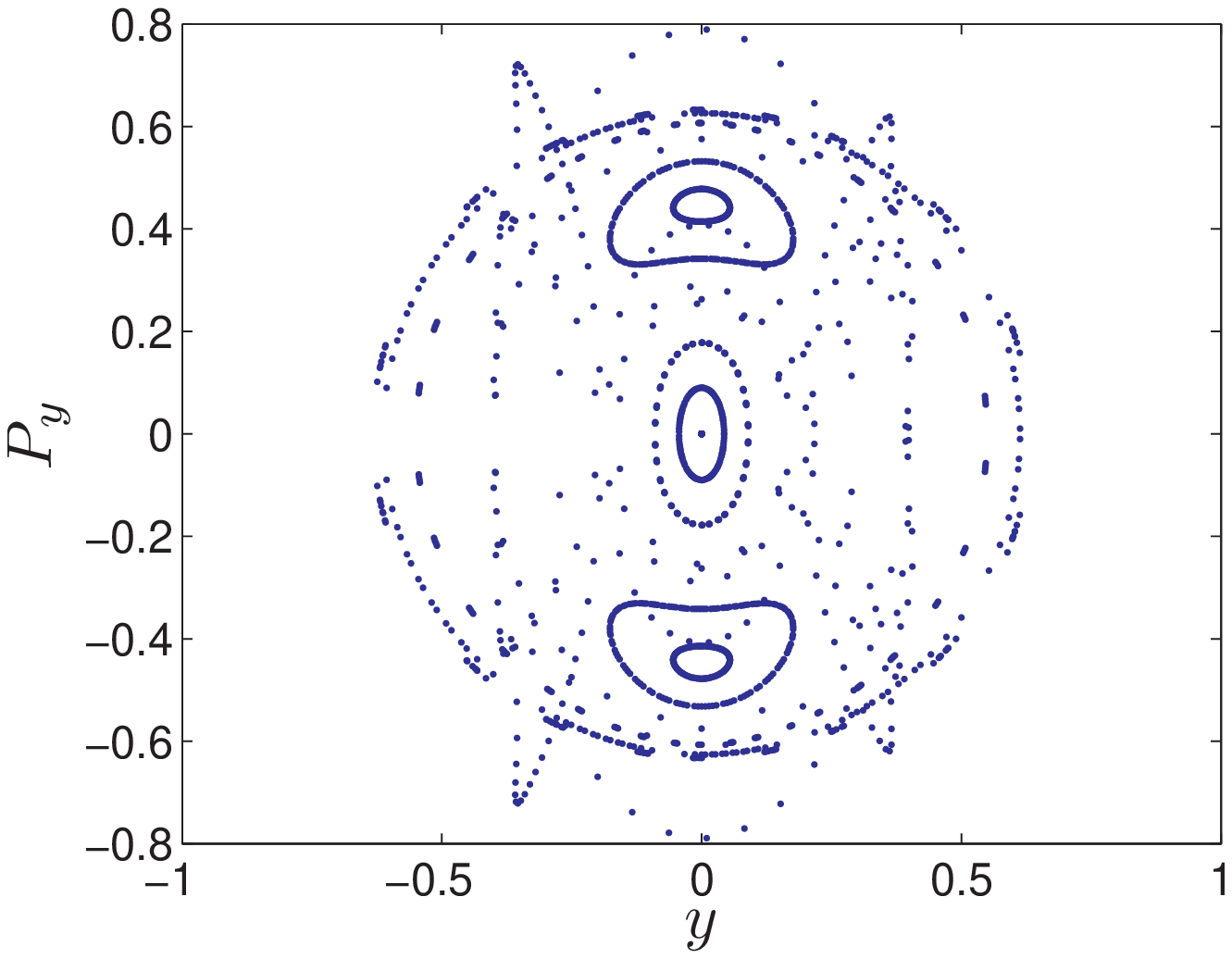}}
  \\
     $(b_{I})$\subfigure{\includegraphics[width=0.25\textwidth]{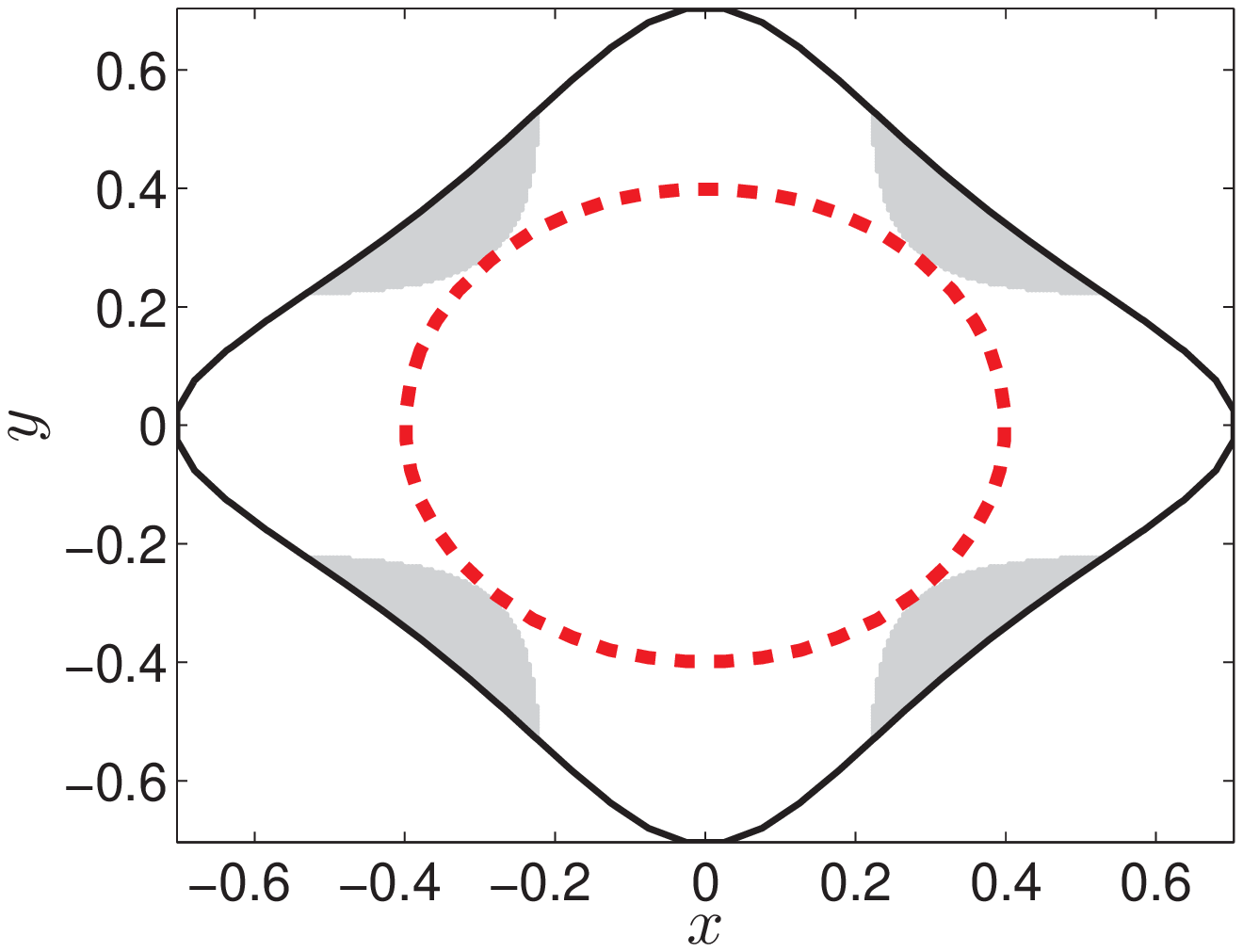}}
   $(b_{II})$\subfigure{\includegraphics[width=0.25\textwidth]{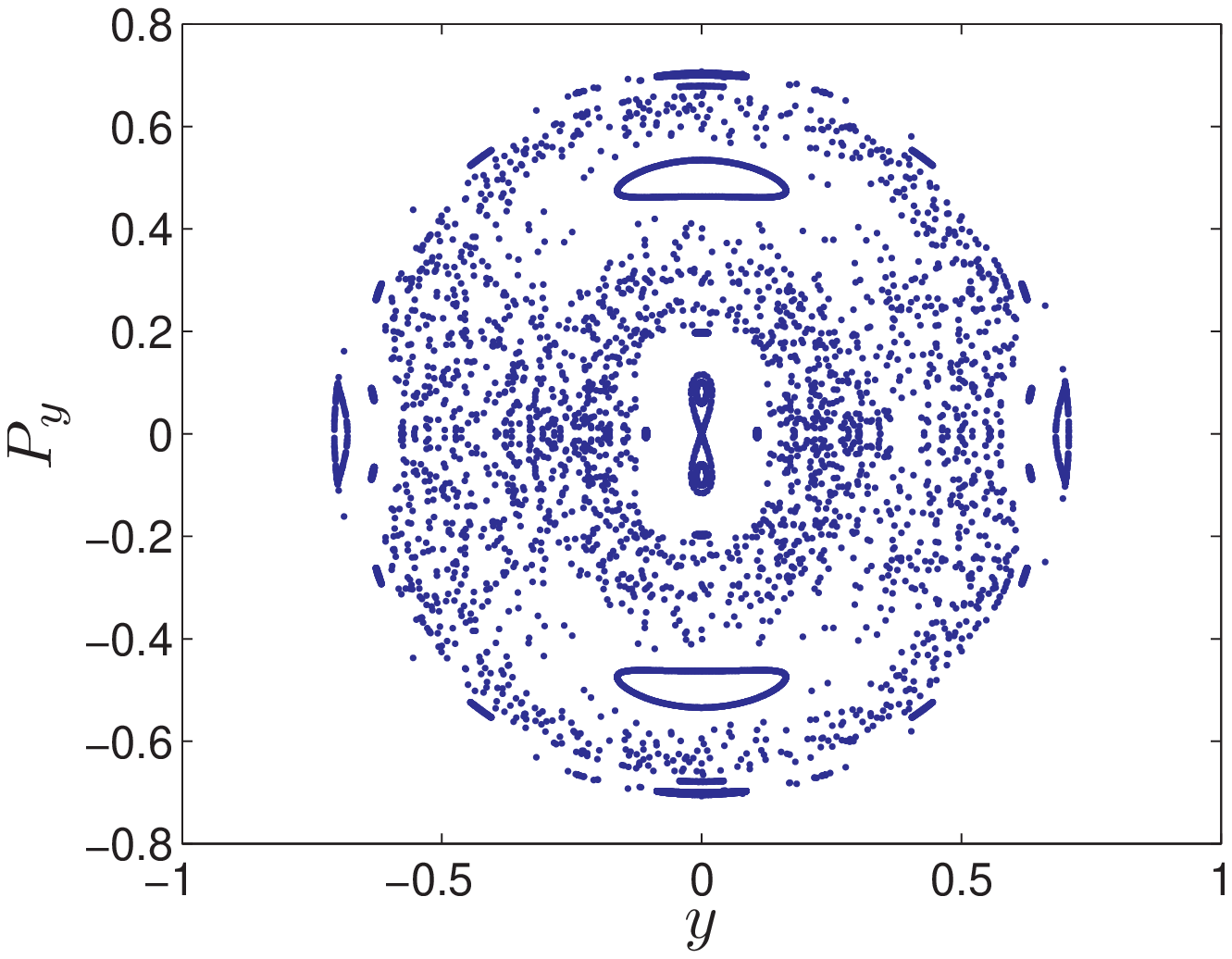}}
  $(b_{III})$\subfigure{\includegraphics[width=0.25\textwidth]{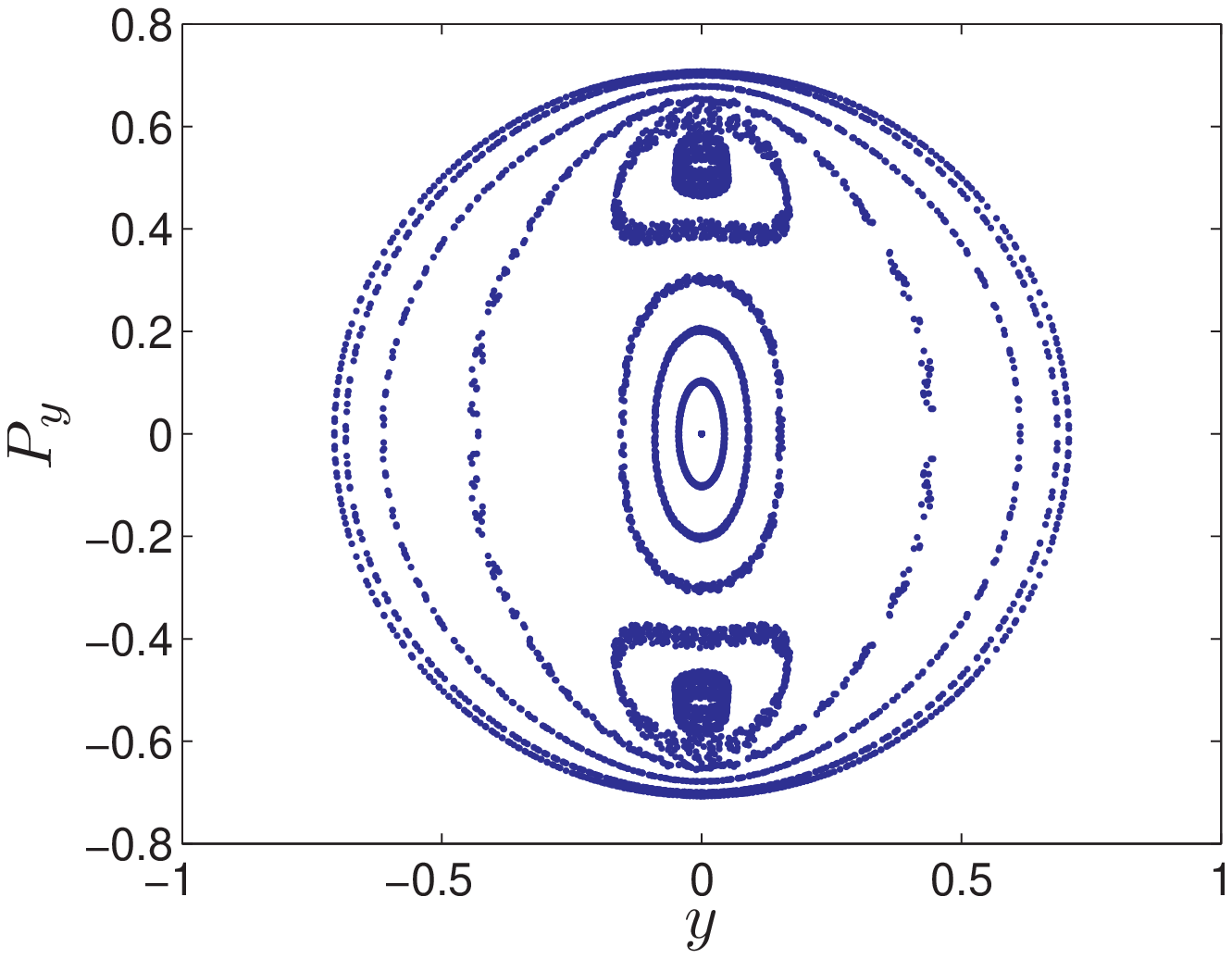}}
  \\
    $(c_{I})$\subfigure{\includegraphics[width=0.25\textwidth]{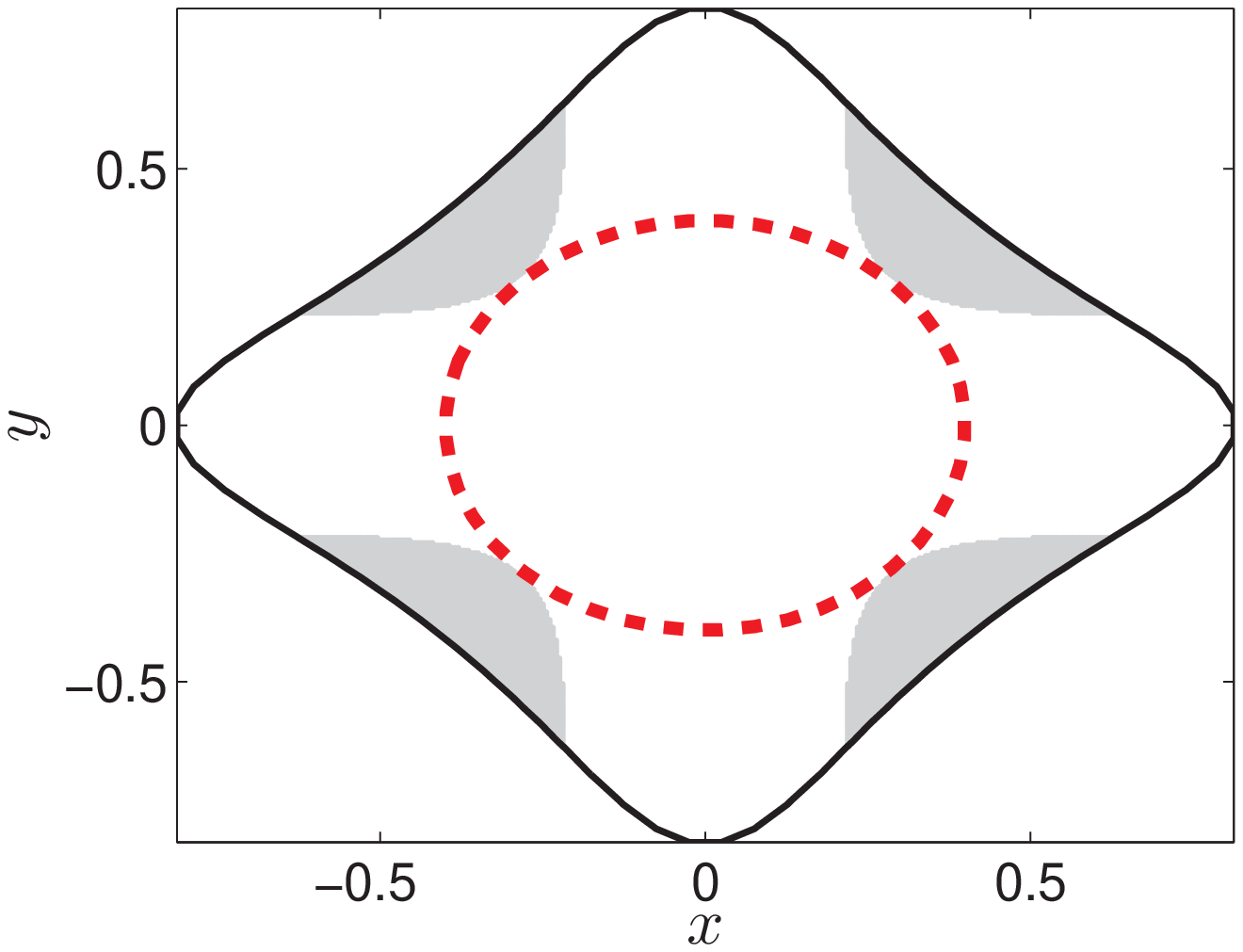}}
   $(c_{II})$\subfigure{\includegraphics[width=0.25\textwidth]{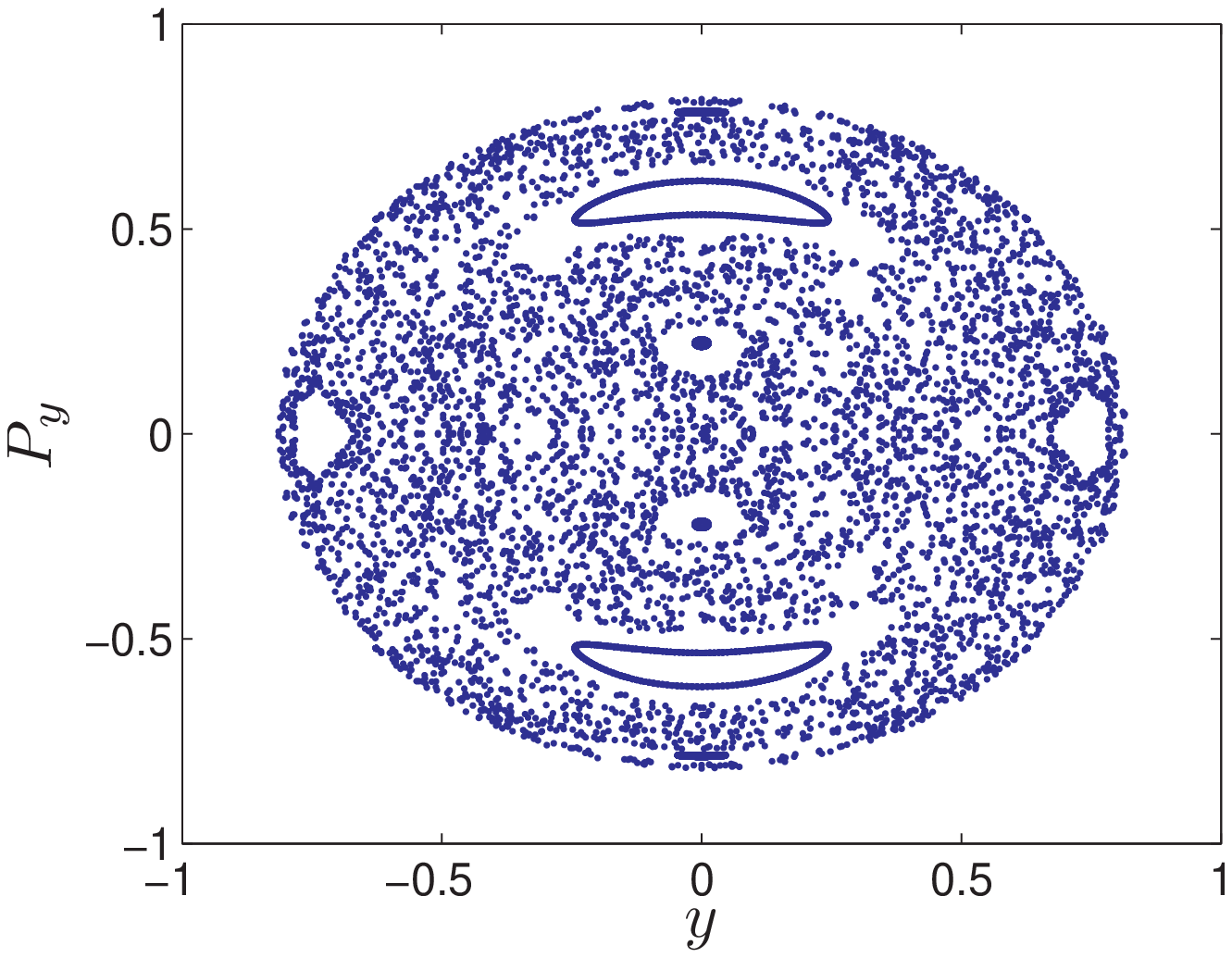}}
   $(c_{III})$\subfigure{\includegraphics[width=0.25\textwidth]{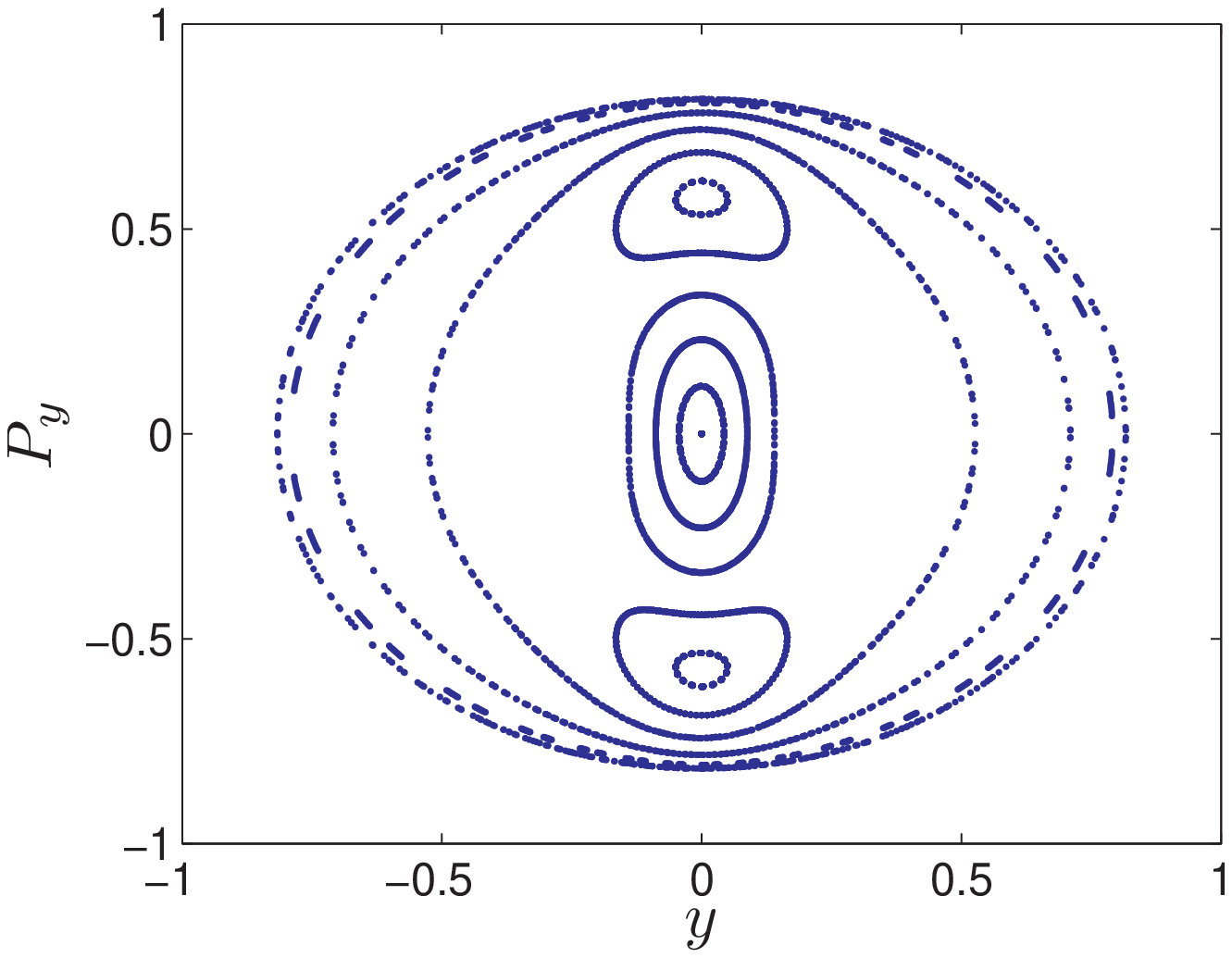}}
\\
    $(d_{I})$\subfigure{\includegraphics[width=0.25\textwidth]{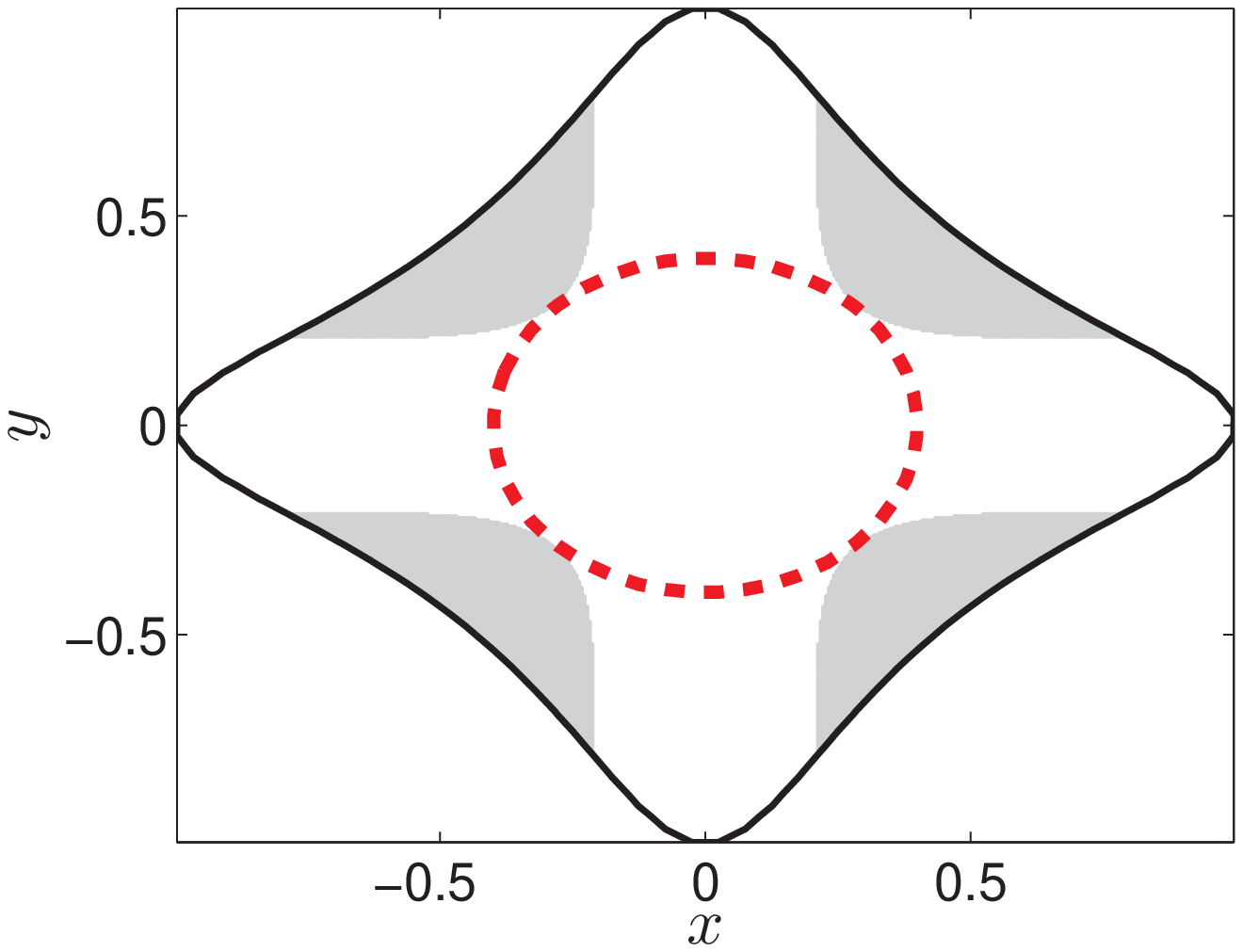}}
   $(d_{II})$\subfigure{\includegraphics[width=0.25\textwidth]{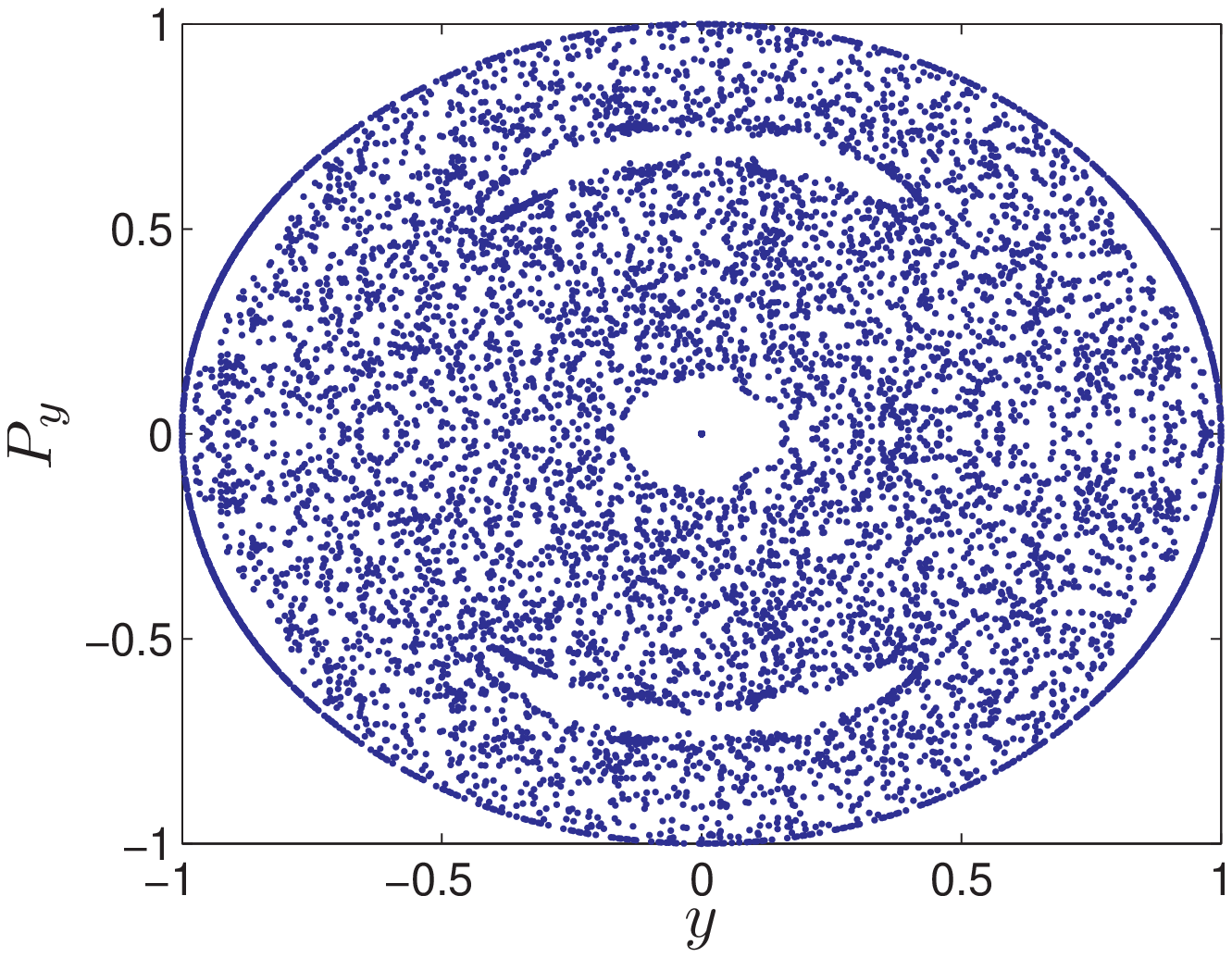}}
   $(d_{III})$\subfigure{\includegraphics[width=0.25\textwidth]{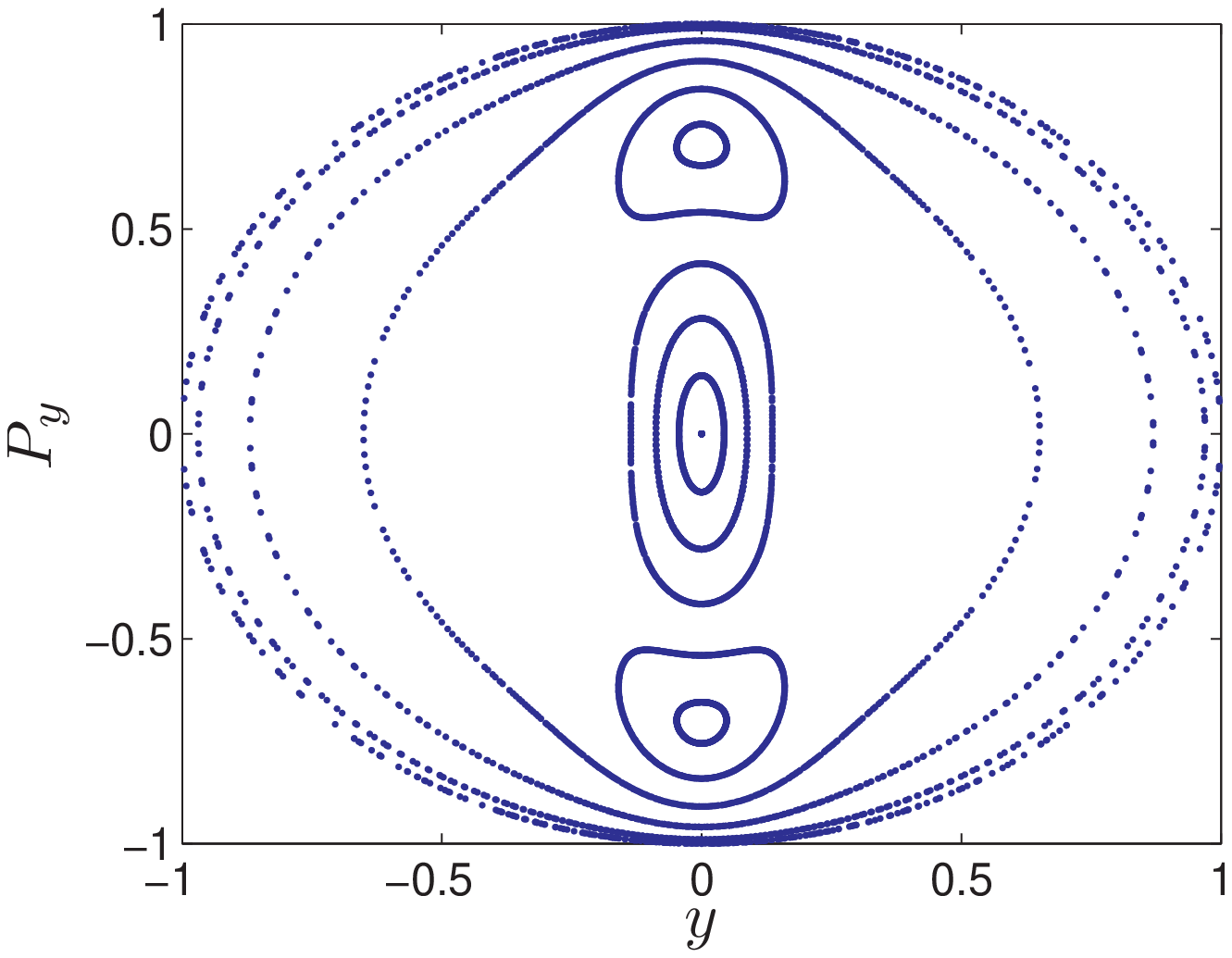}}

  \caption{Effect of control on the model of Eq. \eqref{harmonic}. The first column shows the physical region (closed black curves) corresponding to the sequence of energies,
  $\frac{1}{5},\frac{1}{4},\frac{1}{3},\frac{1}{2}$ corresponding to $a,b,c,d$. The regions of negative eigenvalues (instability) are shown in gray. The second column shows the Poincar\'e
  plots for the uncontrolled system. The third column shows the Poincar\'e plots for the controlled system (Eq. \eqref{harmonicControl} with $r=0.4$ for all cases). The (red) dashed line in column one
  shows the boundary of the control modification.}

\end{figure*}

\begin{figure*}
  \centering
   $(a_{I})$\subfigure{\includegraphics[width=0.25\textwidth]{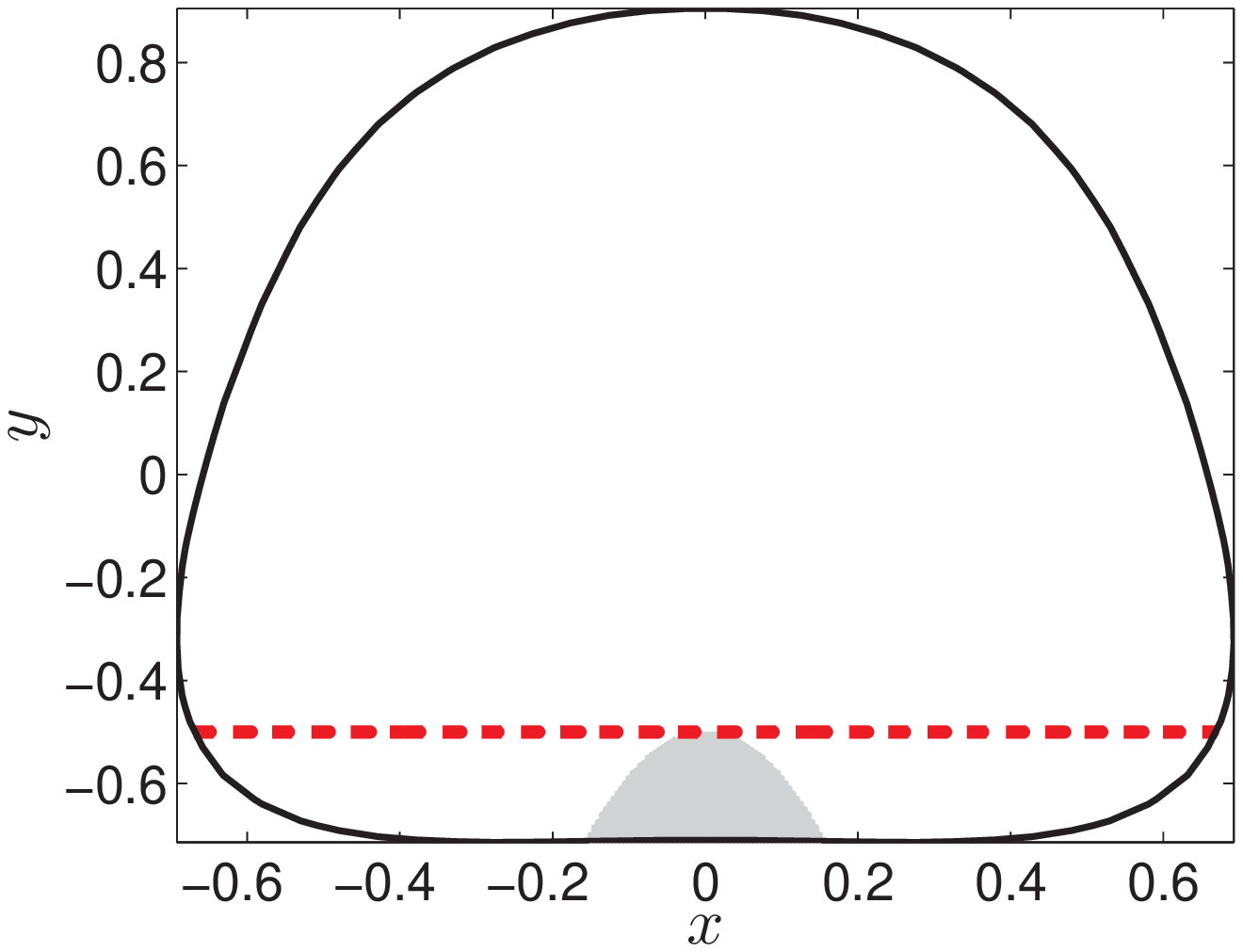}}
   $(a_{II})$\subfigure{\includegraphics[width=0.25\textwidth]{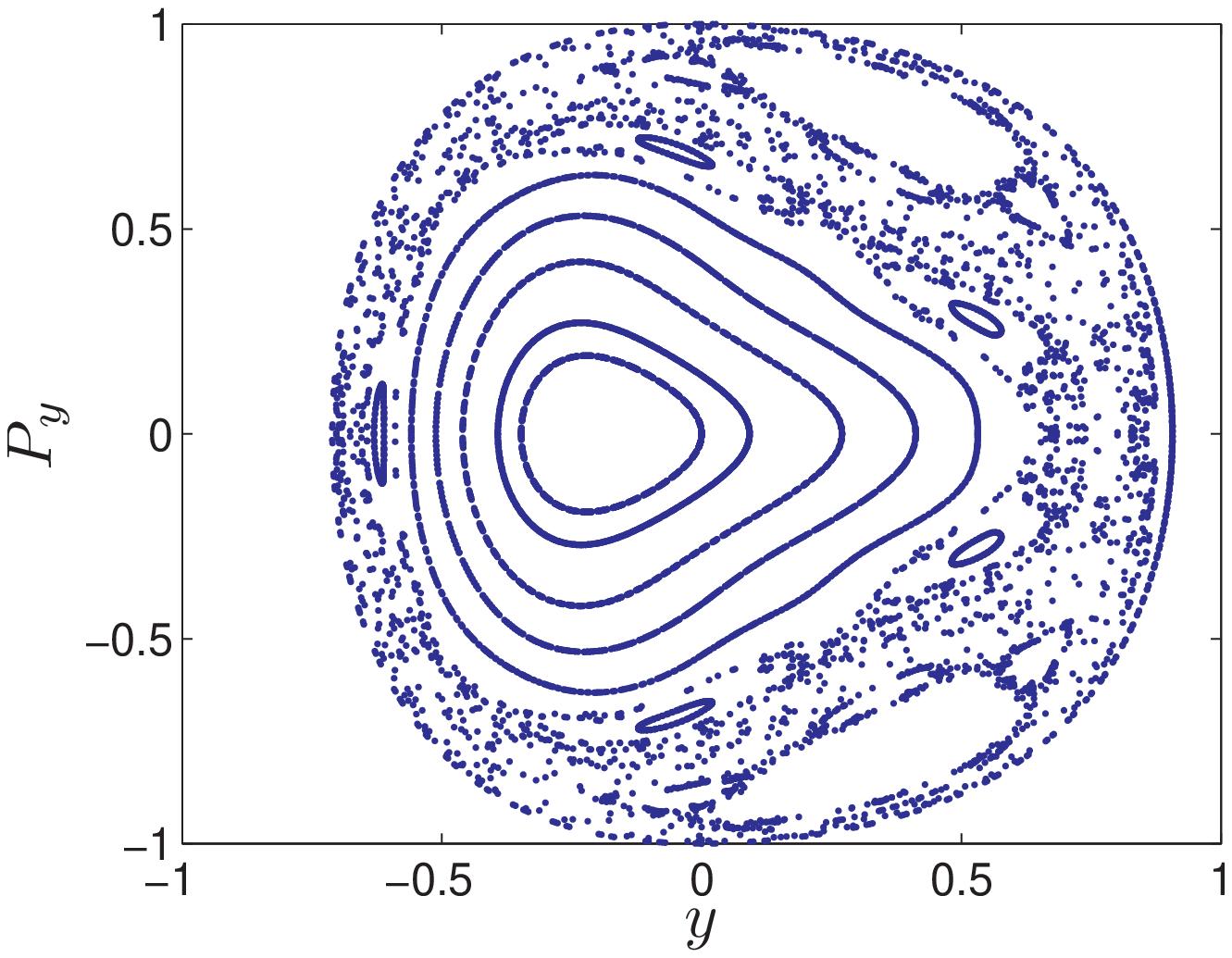}}
   $(a_{III})$\subfigure{\includegraphics[width=0.25\textwidth]{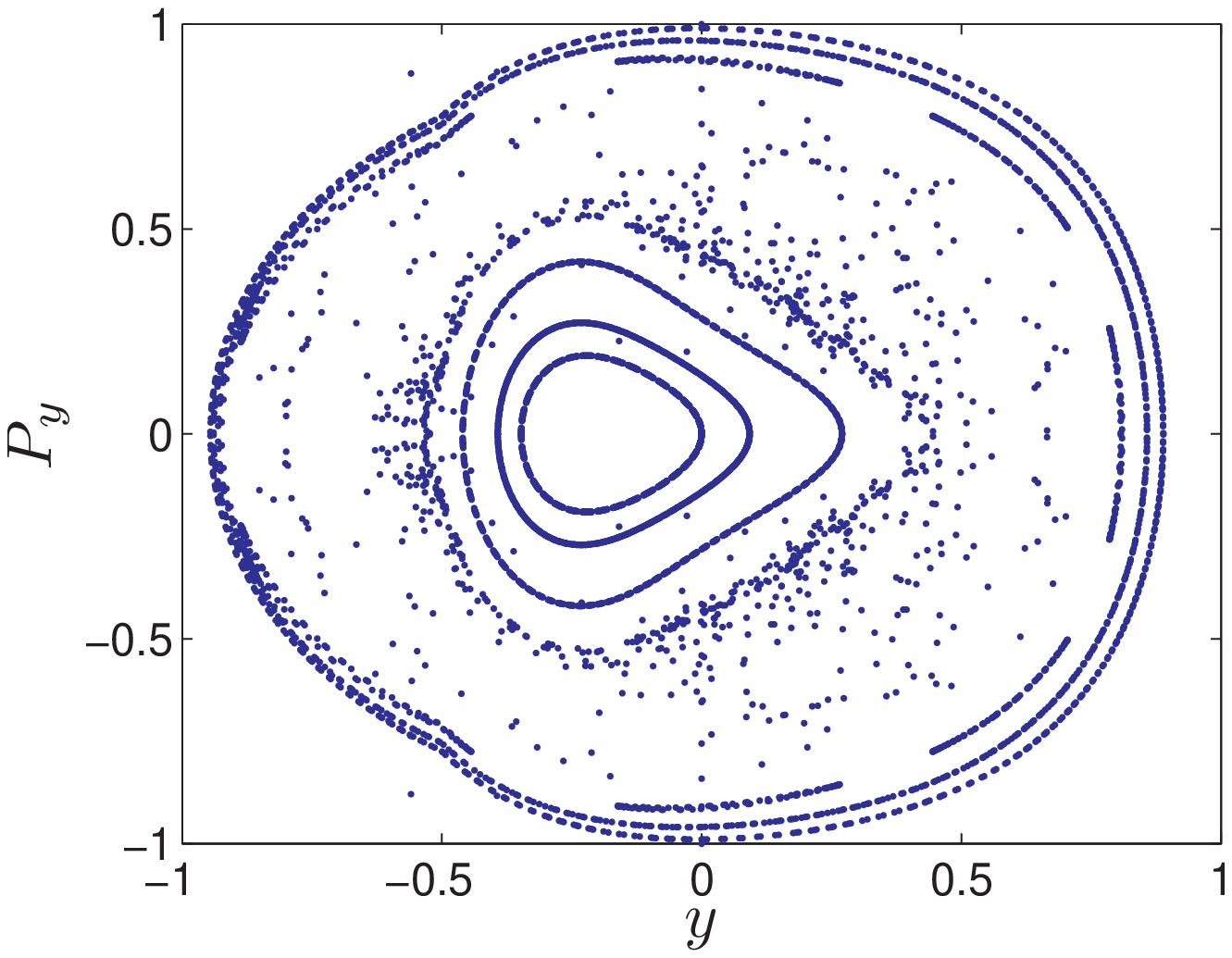}}
\\
   $(b_{I})$\subfigure{\includegraphics[width=0.25\textwidth]{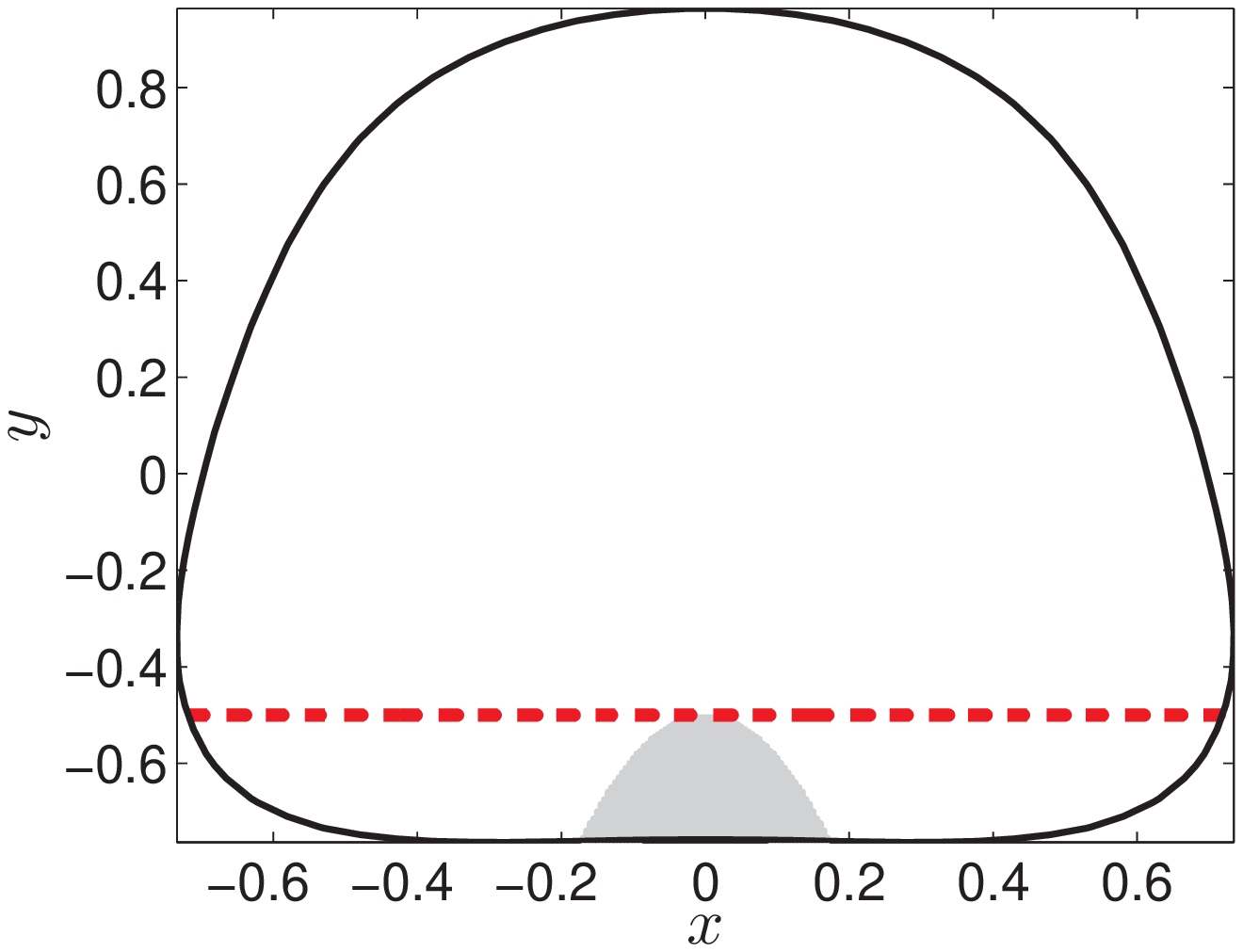}}
   $(b_{II})$\subfigure{\includegraphics[width=0.25\textwidth]{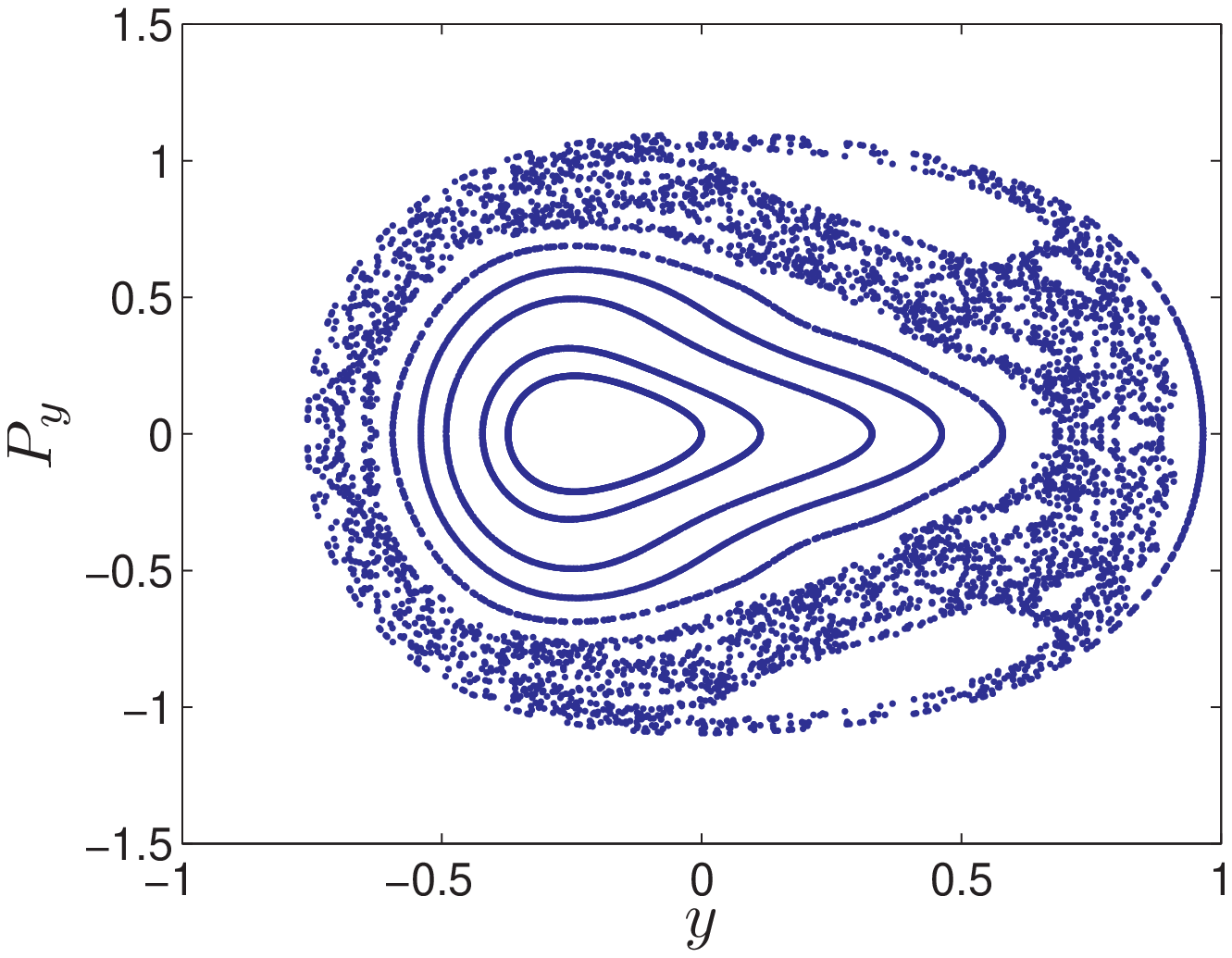}}
   $(b_{III})$\subfigure{\includegraphics[width=0.25\textwidth]{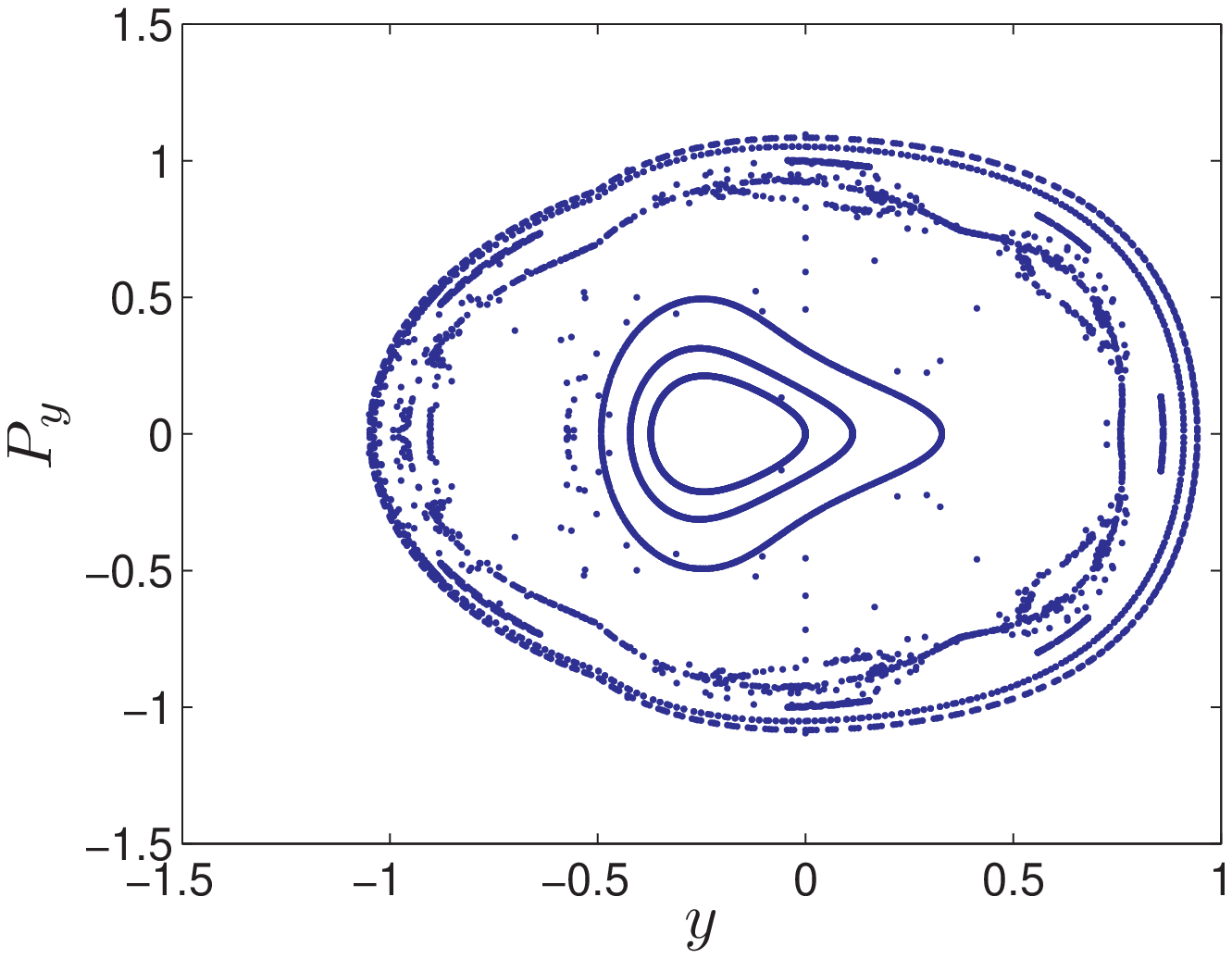}}
  \\
   $(c_{I})$\subfigure{\includegraphics[width=0.25\textwidth]{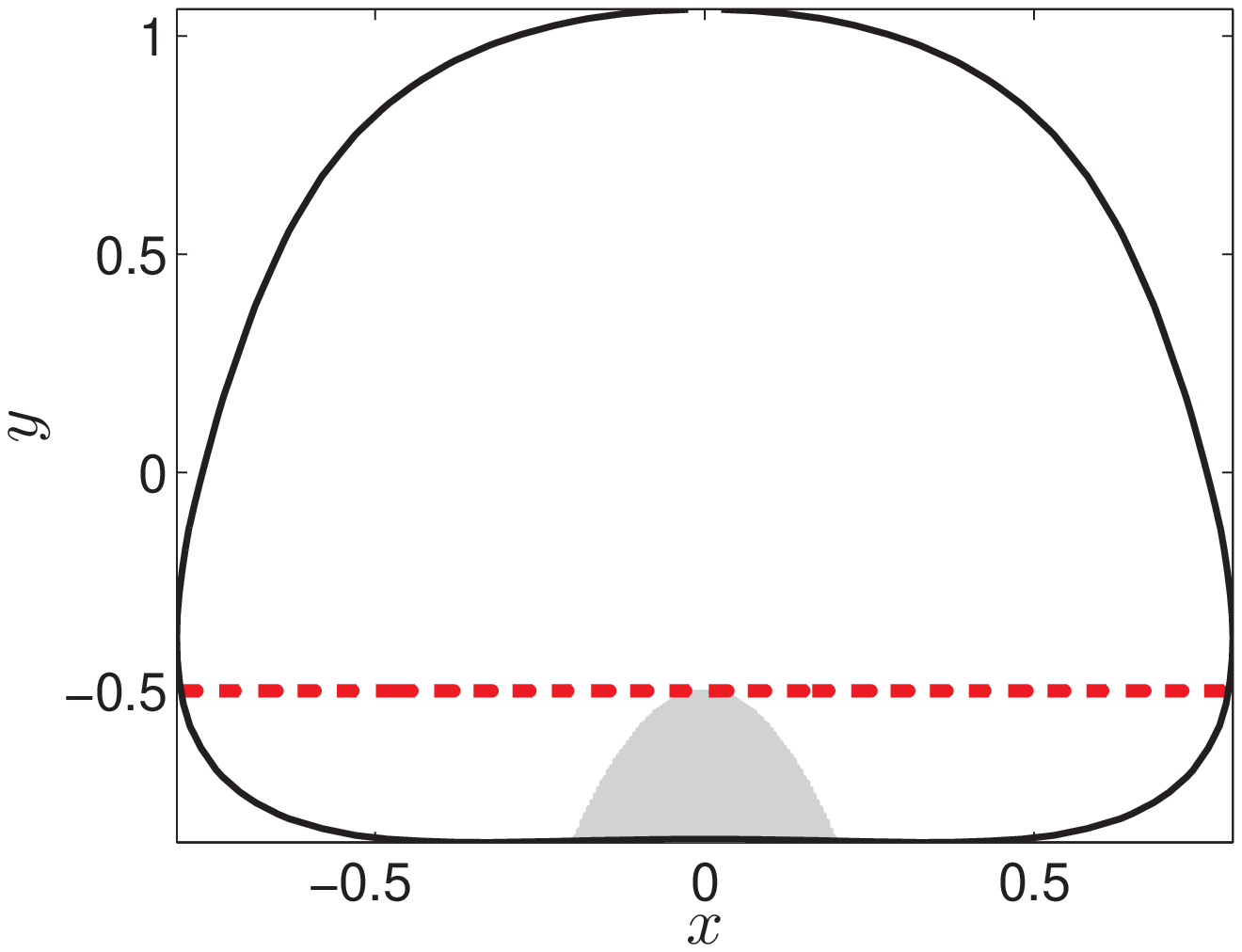}}
   $(c_{II})$\subfigure{\includegraphics[width=0.25\textwidth]{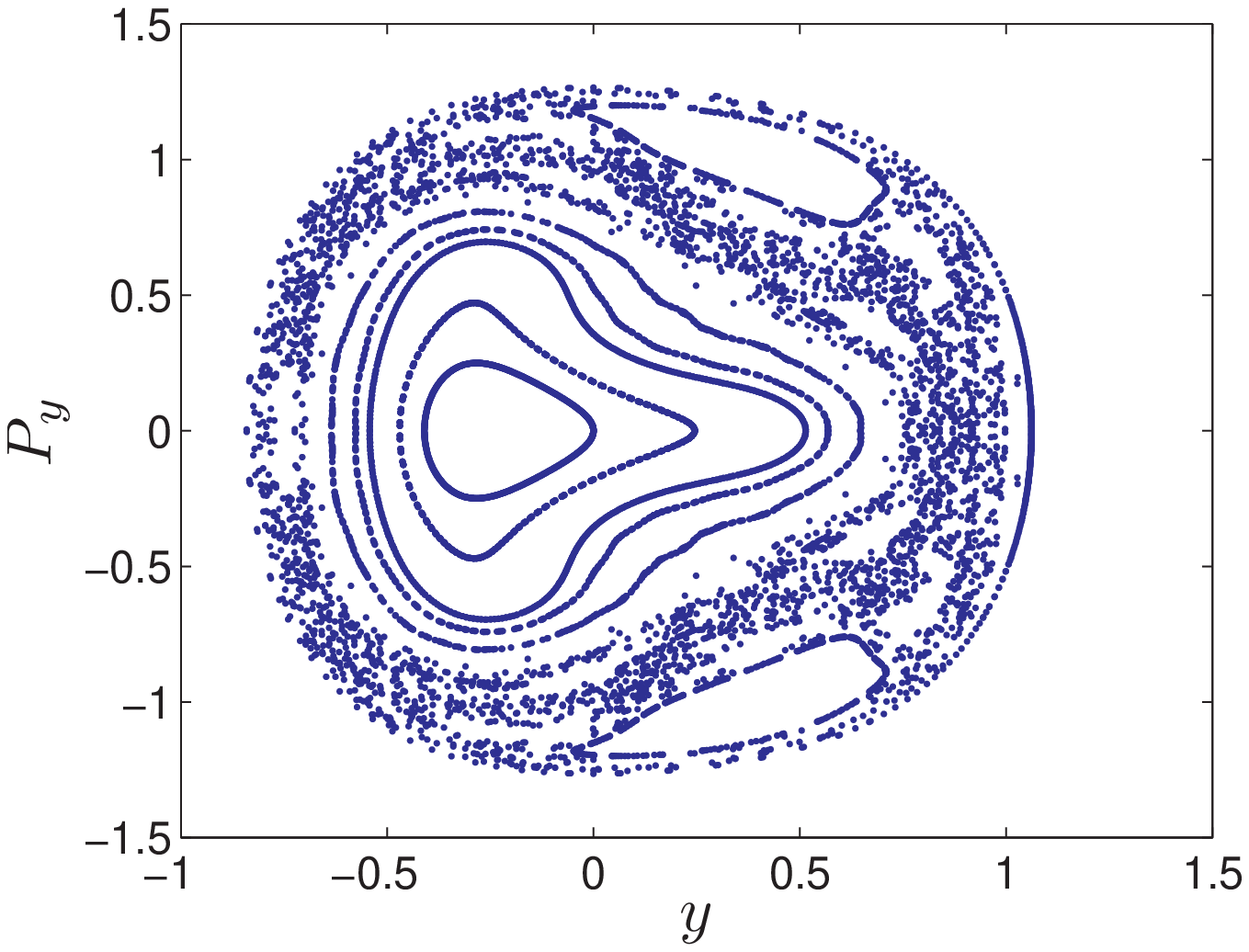}}
   $(c_{III})$\subfigure{\includegraphics[width=0.25\textwidth]{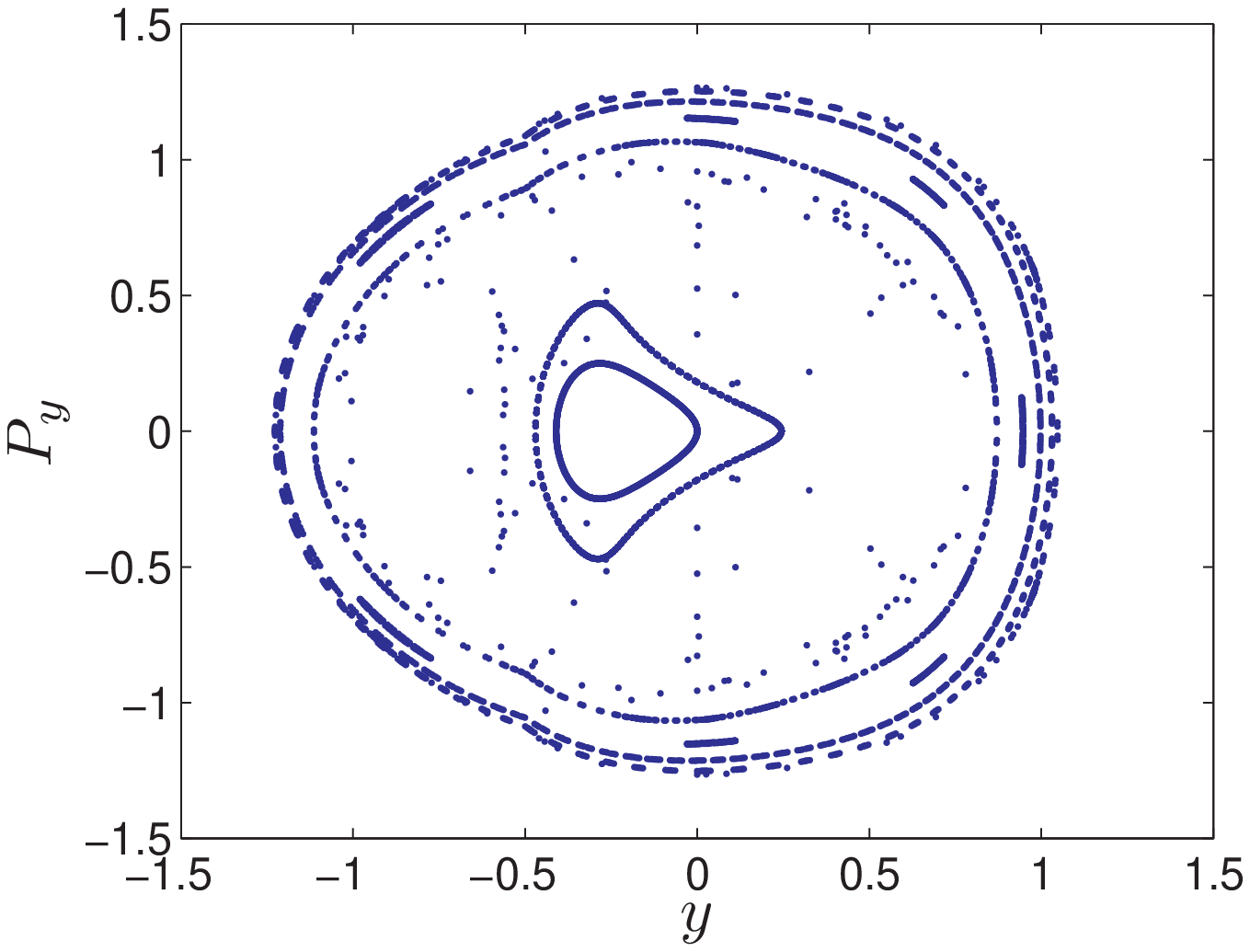}}
  \\
   $(d_{I})$\subfigure{\includegraphics[width=0.25\textwidth]{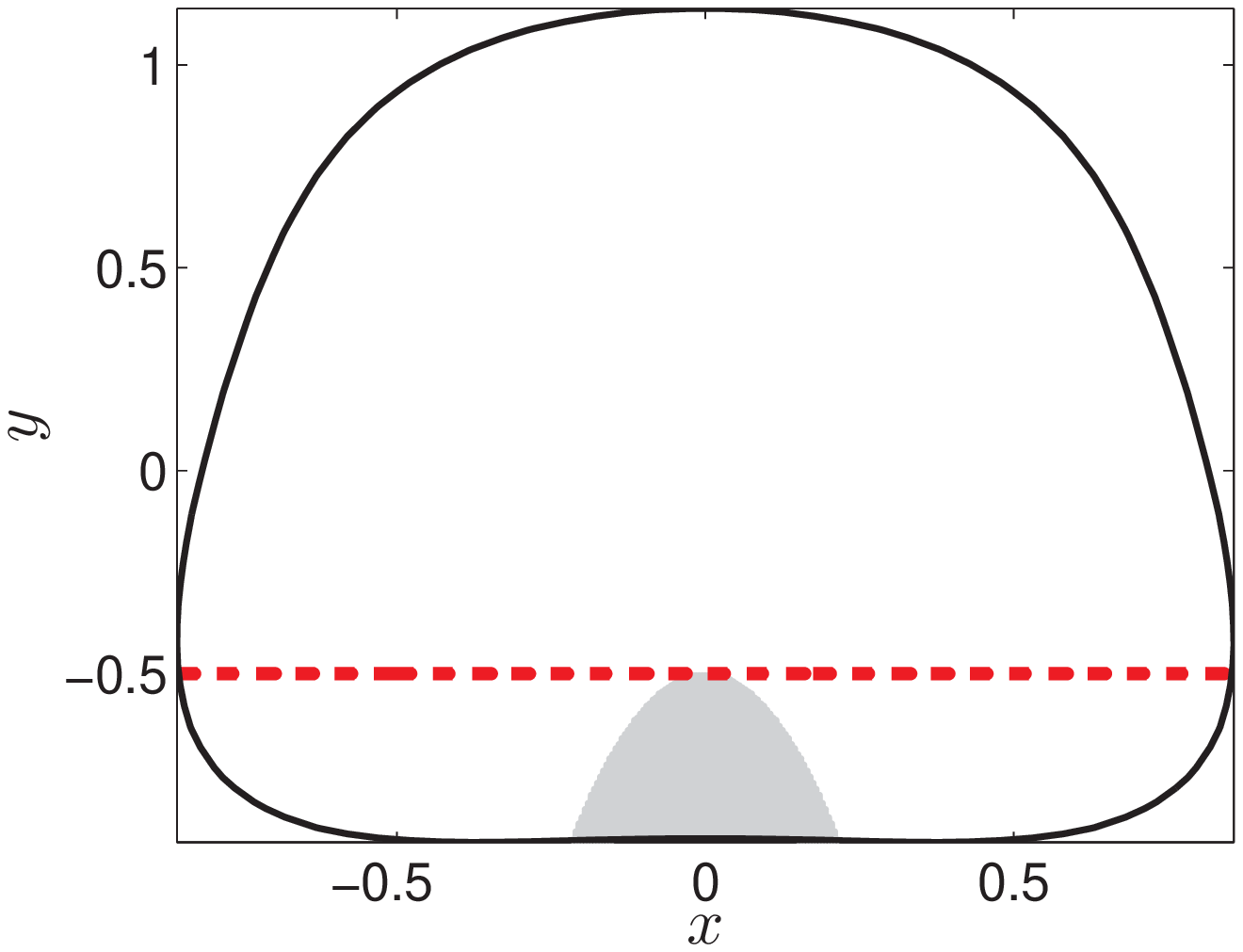}}
   $(d_{II})$\subfigure{\includegraphics[width=0.25\textwidth]{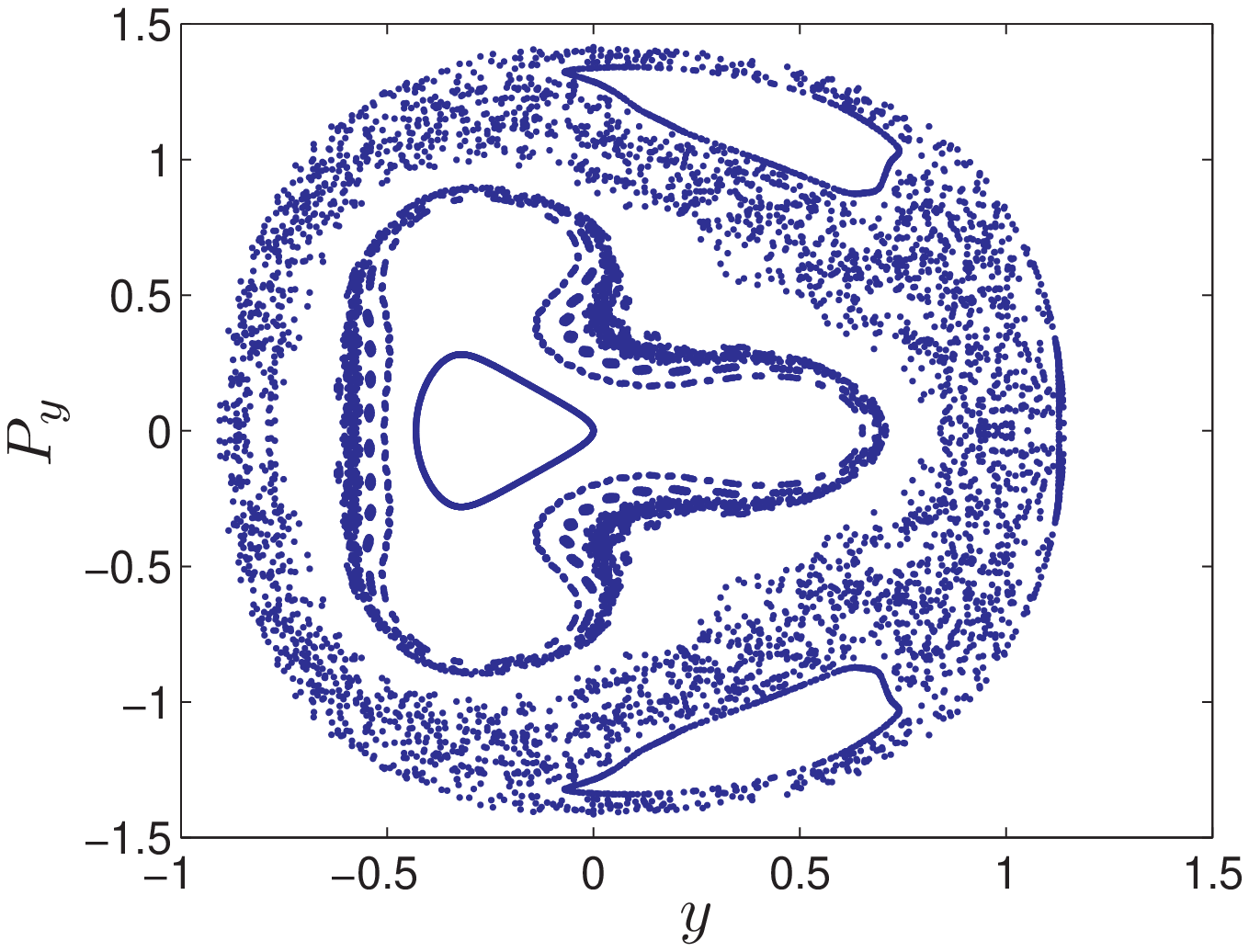}}
   $(d_{III})$\subfigure{\includegraphics[width=0.25\textwidth]{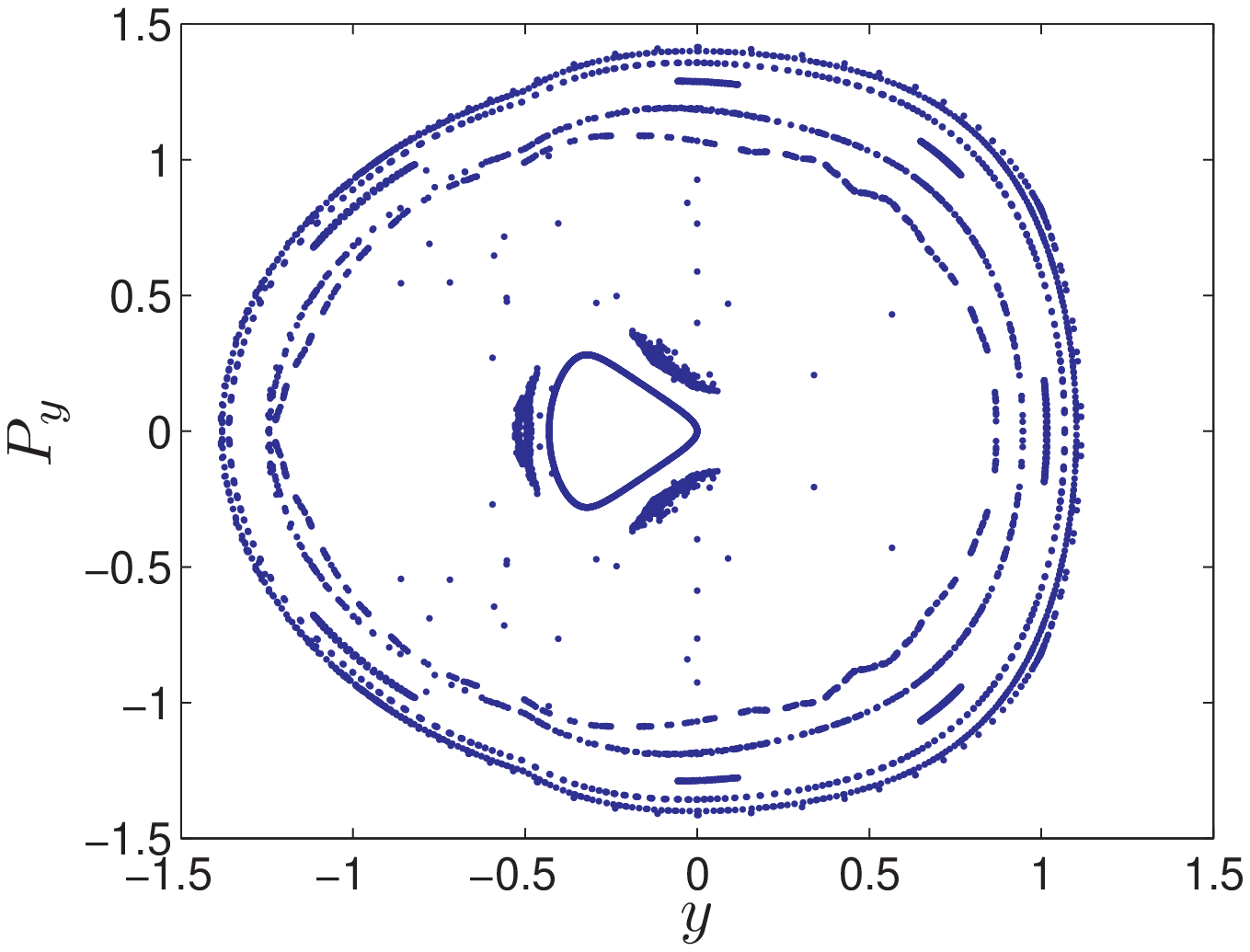}}
  \\
    $(e_{I})$\subfigure{\includegraphics[width=0.25\textwidth]{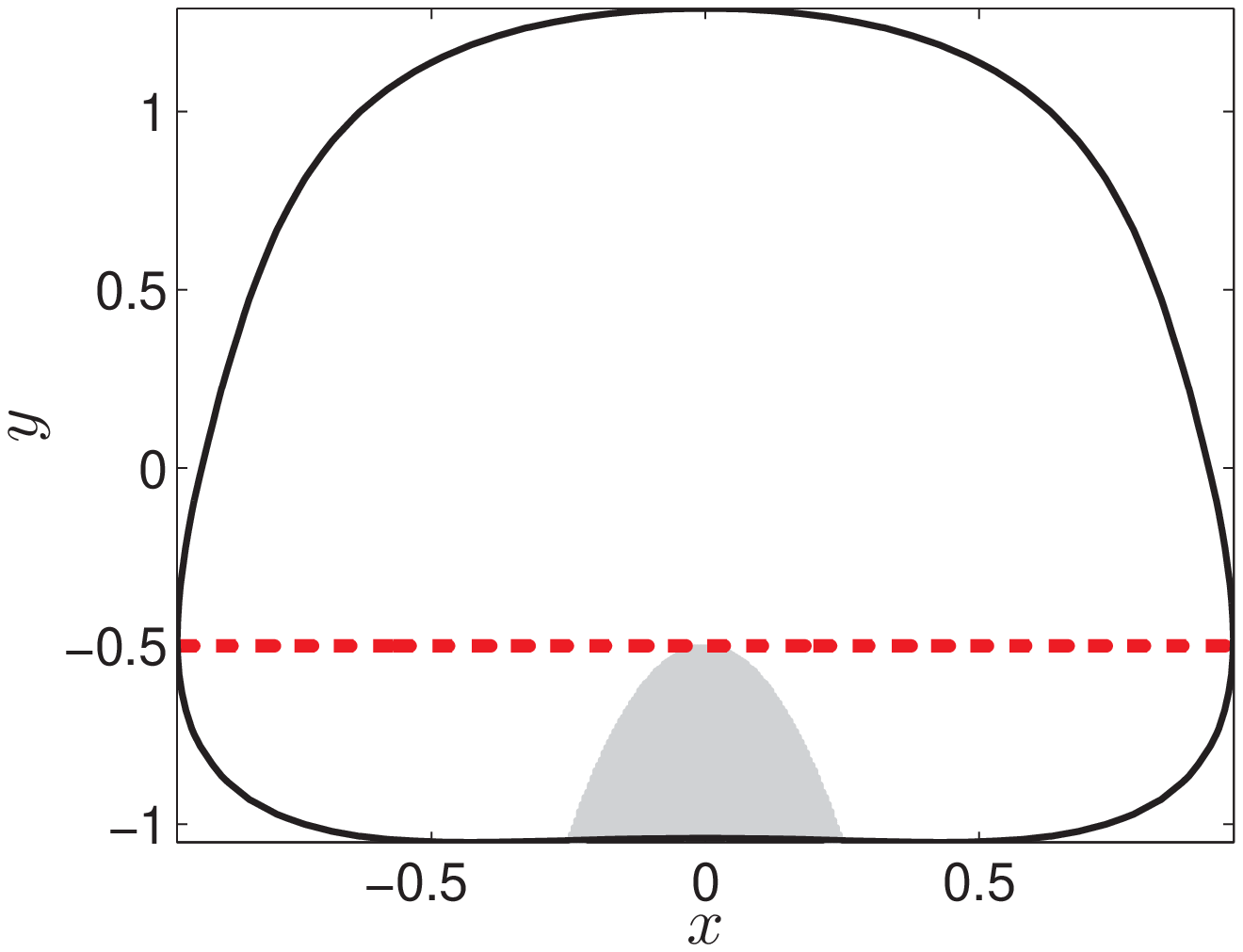}}
   $(e_{II})$\subfigure{\includegraphics[width=0.25\textwidth]{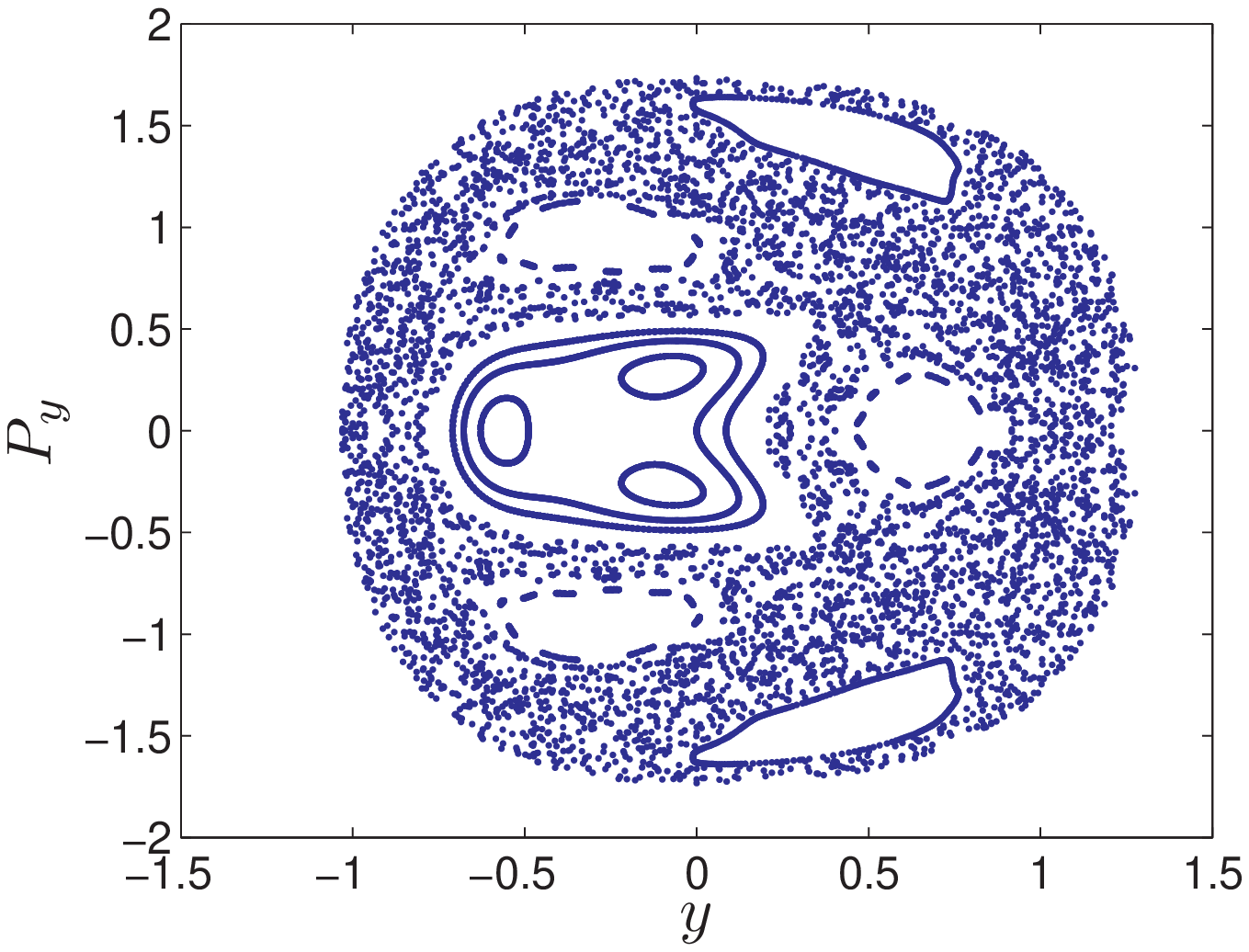}}
   $(e_{III})$\subfigure{\includegraphics[width=0.25\textwidth]{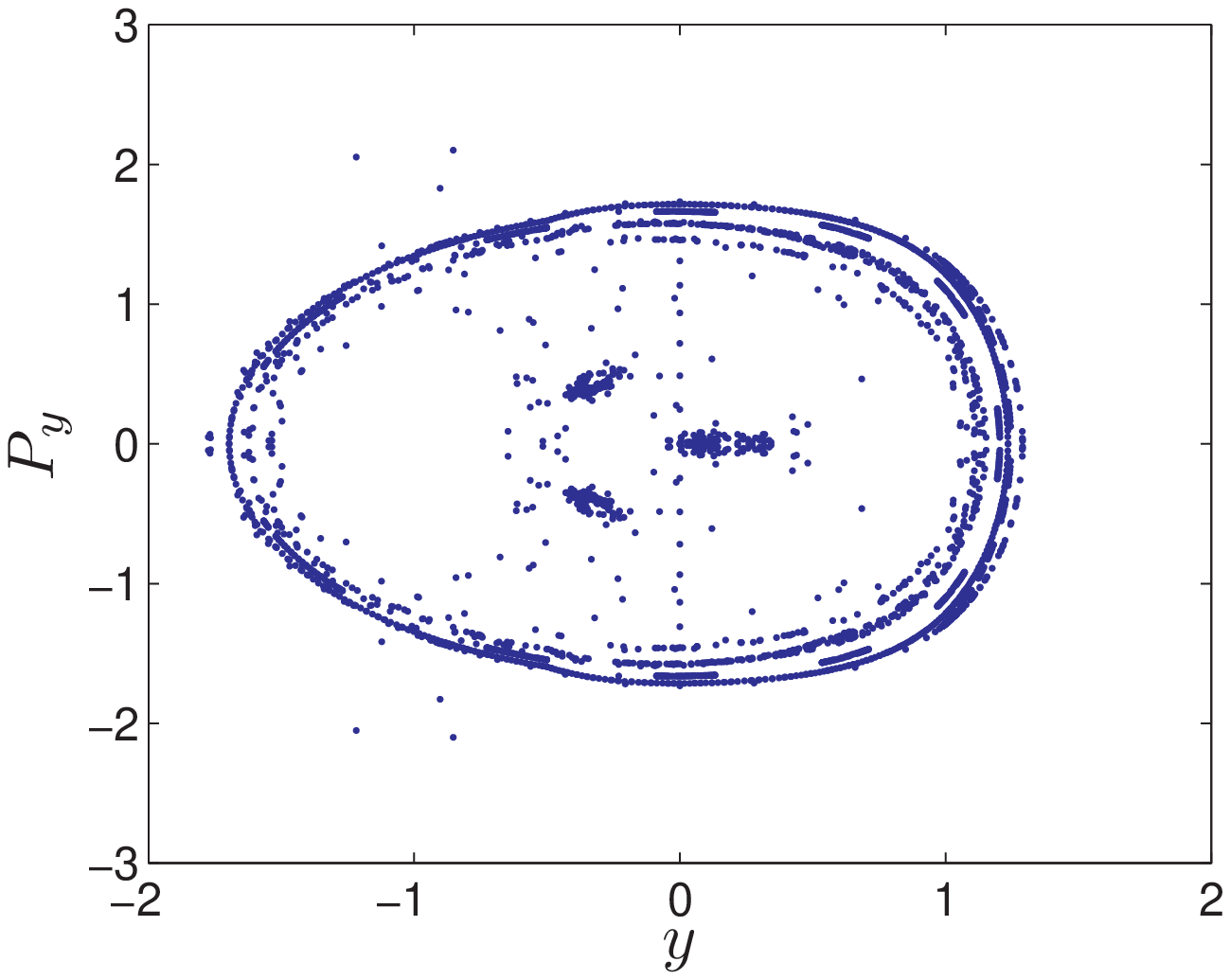}}
  \caption{Effect of control on the model of Eq. \eqref{toda}. The first column shows the physical region (closed black curves) corresponding to the sequence of energies,
  $0.5,0.6,0.8,1.0,1.5$ corresponding to $a,b,c,d,e$. The regions of negative eigenvalues (instability) are shown in gray. The second column shows the Poincar\'e
  plots for the uncontrolled system. The third column shows the Poincar\'e plots for the controlled system (Eq. \eqref{todaControl} with $\alpha=0.5$ for all cases). The (red) dashed line in column one
  shows the boundary of the control modification.}
\end{figure*}

\section{Theory}
\par It has been shown ~\cite{appell,BenZionPRL} that a Hamiltonian system of the
form (we use the summation convention)
\begin{equation}\label{q1}
H= {{p^i}^2 \over 2M} + V(x),
\end{equation}
where $V$ is a function of space variables alone, can be put into
the equivalent form
\begin{equation}\label{q2}
H_G = {1 \over 2M} g_{ij}p^i p^j,
\end{equation}
where $g_{ij}$ is conformal and  is a function of the coordinates
alone. One can easily see that the orbits described by the Hamilton
equations for \eqref{q2} coincide with the geodesics on a Riemannian
space associated with the metric $g_{ij}$, i.e., it follows directly
from the Hamilton equations associated with \eqref{q1}
that~\cite{gutzwiller}
\begin{equation}\label{q3}
{\ddot x}_\ell = -\Gamma_\ell^{mn} {\dot x}_m {\dot x}_n,
\end{equation}
where the connection form $\Gamma_\ell^{mn}$ is given by
\begin{equation}\label{q4}
\Gamma_\ell^{mn} = {1\over 2} g_{\ell k} \bigl\{ {\partial g^{km}
\over
\partial x_n}+  {\partial g^{k n} \over
\partial x_m}- {\partial g^{n m} \over
\partial x_k} \bigr\},
\end{equation}
and $g^{ij}$ is the inverse of $g_{ij}$.
\par  For a metric of conformal form
\begin{equation}\label{q5}
g_{ij} = \varphi \delta_{ij},
\end{equation}
with inverse $g^{ij}= \varphi^{-1}\delta_{ij}$,on the hypersurface
defined by $H_G=H=E=constant$, and assuming that the observable
momenta are the same at every $t$, the requirement of equivalence
implies that ~\cite{BenZionPRL}
\begin{equation}\label{q6}
\varphi = {E \over E-V(x)}.
\end{equation}
\par To see that the Hamilton equations obtained from \eqref{q2} can,
be put into correspondence with those obtained from the Hamiltonian
of the potential model \eqref{q1}, we first note, from the Hamilton
equations for \eqref{q2}, that
\begin{equation}\label{q7}
{\dot x}_i = {\partial H_G \over \partial p^i}= {1 \over M}g_{ij}
 p^j.
\end{equation}
We then define the {\it velocity field}
\begin{equation}\label{q8}
{\dot x}^j \equiv  g^{ji}{\dot x}_i = {1 \over M}p^j,
\end{equation}
coinciding formally with one of the Hamilton equations implied by
\eqref{q1}, for which we label the coordinates ${x^j}$.
\par To complete our correspondence with the dynamics induced by
\eqref{q1}, consider the Hamilton equation for ${\dot p}^i$
generated by $H_G$,
\begin{equation}\label{q9}
{\dot p}^\ell = -{\partial H_G \over \partial  x_\ell} = -{1 \over
2M}{\partial g_{ij} \over
\partial x_\ell}  p^i p^j .
\end{equation}
With \eqref{q8} this becomes
\begin{equation}\label{q16}
 {\ddot x}^\ell = -M^\ell_{mn}{\dot x}^m {\dot x}^n,
\end{equation}
where
\begin{equation}\label{q17}
M^\ell_{mn}\equiv {1 \over 2}g^{\ell k}{\partial g_{n m} \over
\partial x^k}.
\end{equation}
\par Eq. \eqref{q16} has the form of a geodesic equation, with a
truncated connection form~\cite{BenZionPRL}. Note that performing
parallel transport on the local flat tangent space of the Gutzwiller
manifold (for which $\Gamma_\ell^{mn}$ and $g_{ij}$ are compatible),
the resulting connection, after
 raising the tensor index to reach the Hamilton manifold, is
 exactly the ``truncated'' connection \eqref{q17}.
  \par  Substituting \eqref{q5} and \eqref{q6} into \eqref{q16} and \eqref{q17}, the Kronecker
deltas identify the indices of ${\dot x}^m$ and ${\dot x}^n$; the
resulting square of the velocity cancels a factor of $(E-V)^{-1}$,
leaving the Hamilton-Newton law derived from the Hamilton equations
directly from \eqref{q1}. Eq. \eqref{q16} is therefore a
geometrically covariant form of the Hamilton-Newton law, exhibiting
what can be considered an underlying geometry of standard
Hamiltonian motion.
\par Since the coefficients $M^\ell_{mn}$ constitute a connection
form, they can be used to construct a covariant derivative. It is
this
 covariant derivative which must be used to
compute the rate of transport of the geodesic deviation $\xi^\ell =
x'^\ell - x^\ell$ along
 the (approximately common) motion of neighboring orbits in the
 Hamilton manifold, since it follows the geometrical structure of the
geodesic curves on $\{x^\ell\}$.

\par The relation
\begin{equation}\label{q18}
{\ddot \xi}^\ell =  -2 M^\ell_{mn}{\dot x}^m {\dot \xi}^n -
{\partial M^\ell_{mn}\over \partial x^q} {\dot x}^m{\dot x}^n \xi^q,
\end{equation}
 obtained from \eqref{q16}, can be
factorized in terms of the covariant derivative
\begin{equation}\label{q19}
\xi^\ell_{;n} = {\partial \xi^\ell \over \partial x^n} +
M^\ell_{nm}\xi^m.
\end{equation}
One obtains
\begin{equation}\label{q20}
{{D_M}^2 \over {D_M} t^2} \xi^\ell  = {{R_M}^\ell}_{qmn} {\dot
x}^q{\dot x}^n \xi^m,
\end{equation}
where the index $M$ refers to the connection \eqref{q17}, and
\begin{equation}\label{q21}
{{{R_M} ^\ell}_{qmn}  = {\partial M^\ell_{qm} \over \partial x^n}
-{\partial M^\ell_{qn} \over \partial x^m} + M^k_{qm}M^\ell_{nk} -
M^k_{qn}M^\ell_{mk}}
\end{equation}
corresponds to the curvature associated with the connection form
$M^\ell_{mn}$
 This
expression does not coincide with the curvature of the Gutzwiller
manifold (given by this formula with $\Gamma_{qm}^\ell$ in place of
$M_{qm}^\ell$), but is a {\it dynamical curvature} which is
appropriate for geodesic motion in $\{x^\ell\}$
\par With the conformal metric in noncovariant form \eqref{q5},\eqref{q6} (in the
coordinate system in which \eqref{q6} is defined), the dynamical
curvature \eqref{q21} can be written in terms of derivatives of the
potential $V$, and the geodesic deviation equation \eqref{q20}
becomes
\begin{equation}\label{q22}
{D^2{\bf \xi} \over Dt^2} = - {\cal V}P{\bf \xi},
\end{equation}
where the matrix ${\cal V}$ is given by
\begin{equation}\label{q23}
{\cal V}_{\ell i} =  \bigl\{ {3 \over M^2v^2} {\partial V \over
\partial x^\ell} {\partial V \over \partial x^i} + {1 \over M}
{\partial^2 V \over \partial x^\ell \partial x^i} \bigr\}.
\end{equation}
and
\begin{equation}\label{q24}
P^{ij} = \delta^{ij} - {v^i v^j
 \over v^2},
 \end{equation}
with $ v^i \equiv {\dot x}^i$, defining a projection into a
direction orthogonal to $v^i$.

\par Instability should occur if at least one of the eigenvalues of
 $P{\cal V}P$ is negative,
 in terms of the second covariant derivatives of the
transverse component of the geodesic deviation. This condition is
 easily seen to be equivalent to the same condition imposed on the
 spectrum of ${\cal V}$, and is thus independent of the direction of
 the motion on the orbit.
\par  The condition implied by the geodesic deviation
 equation \eqref{q22}, in terms of
covariant derivatives, in which the orbits are viewed geometrically
as geodesic motion, is a  new condition for instability
~\cite{BenZionPRL}, based on the underlying geometry, for a
Hamiltonian system of the form \eqref{q1}, providing new insight
into the structure of the unstable and chaotic behavior of
Hamiltonian dynamical systems. The method has been successfully
applied to many potential
models~\cite{BenZionPRL,BenZionPRE1,BenZionPRE2,yahalom}. We now
apply the method, which is simple and straightforward, since the
instability criterion is defined locally in coordinate space, to
extract from a chaotic Hamiltonian the part of the dynamics in the
regions for which ${\cal V}$ has negative eigenvalues which lead to
unstable motion .  We show by direct simulation that this procedure
is remarkably effective.
\par The results confirm the notion of the locality of our criterion,
since the (local) removal of parts of the Hamiltonian inducing
instability by this criterion strongly affects the global stability
of the motion.

\section{Results and discussion}

\par In the following we give some examples of Hamiltonian motion
which are unstable and exhibit chaotic behavior.  In two dimensions,
the terms in the potential that break rotational symmetry have a
form, in these examples, that induces chaotic behavior. The matrix
${\cal V}$ computed over the physically accessible region exhibits
certain regions with negative eigenvalues. We have arranged our
simulation to replace the Hamiltonian {\it in the regions of
negative eigenvalues} by a Hamiltonian in which the symmetry
breaking terms are either removed or for which the coupling
coefficient is decreased, but in regions of positive eigenvalues,
the Hamiltonian remains in its original form. The changes are
carried in relatively small regions of the configuration space in
the neighborhood of the regions of negative eigenvalues.

We take for illustration here a simple and important case of coupled
harmonic oscillators with perturbation
\begin{equation}\label{harmonic}
V(x,y)=\frac{1}{2}(x^2+y^2)+6x^2y^2
\end{equation}
and a generalization of the Toda potential
\begin{equation}\label{toda}
V(x,y)=\frac{1}{2}(x^2+y^2)+x^2y-\frac{1}{3}y^3+\frac{3}{2}x^4+\frac{1}{2}y^4.
\end{equation}

The first of these is known to generate chaotic behavior; as we have
shown previously, the transition to chaotic behavior as a function
of energy and of the coupling to the perturbation is well described
by our geometric criterion \cite{BenZionPRL}. In the present work,
we set the coupling to the chaos inducing perturbation to zero in
the regions for which the geometric criterion results in negative
eigenvalues.

In our previous study of the second example Eq. \eqref{toda} we
studied the sensitivity of the dynamical behavior to the choice of
energy \cite{BenZionPRE2}. In the present work, at the energies for
which the system exhibits chaotic behavior we have removed the chaos
inducing perturbation in the regions for which the geometric
criterion results in negative eigenvalues.

Fig. 1 shows the effects on the dynamical behavior of the change of
the coupling to a stable value in the regions of negative
eigenvalues for the perturbed oscillator potential \eqref{harmonic}.
The results for different values of energy are shown; these values
correspond to different sizes of the physically accessible region.
Note that the radius of the region of positive eigenvalues does not
change in this example.

The first row shows the physical region and the location of negative
eigenvalues. The interior of the circle corresponds to a region in
which no negative eigenvalue occur, chosen in this way to simplify
the computation. The second row shows the surface of section
Poincar\'e plot(surface of section) of the original uncontrolled
Hamiltonian indicating chaotic dynamics. The third row shows the
Poincar\'e plot generated by the controlled potential:
\begin{equation}\label{harmonicControl}
V(x,y)=\left\{\begin{matrix}
 \frac{1}{2}(x^2+y^2)+6x^2y^2 & x^2+y^2<r^2\\
 \frac{1}{2}(x^2+y^2) & x^2+y^2\geq r^2
\end{matrix}\right.
\end{equation}

where $r$ stands for the radius of the region of positive
eigenvalues.

Fig. 2 shows the effect of control on second system \eqref{toda},
using the controlled potential:
\begin{equation}\label{todaControl}
V(x,y)=\left\{\begin{matrix}
 \frac{1}{2}(x^2+y^2)+x^2y-\frac{1}{3}y^3+\frac{3}{2}x^4+\frac{1}{2}y^4 & y > -\alpha\\
 \frac{1}{2}(x^2+y^2) & y \leq  -\alpha
\end{matrix}\right.
\end{equation}
where $\alpha$ stands for the limit (a horizontal dashed line) of
the region of positive eigenvalues.

In both cases, for sufficiently high energies the uncontrolled
systems become less stable and a chaotic signature appears. We
examine the method for different values of the energy, all
corresponding to chaotic motion of the uncontrolled system. One can
easily see that the Poincar\'e plots in controlled systems present
almost completely regular motion.

\section{Conclusion}

 We see from these computations that in regions of positive eigenvalues
the Hamiltonian with chaos inducing terms do not generate
instability. If the chaos inducing terms are removed or decreased in
coupling in the regions of negative eigenvalues, the chaotic motion
of the system is reduced or disappears. We conclude that the chaotic
behavior of the system is associated with local properties of the
Hamiltonian, and that our local criterion is effective in
identifying these regions. Furthermore, our procedure provides an
effective method for control of chaotic systems of this type.

\end{document}